\documentclass[11pt,draftclsnofoot,onecolumn,journal,letterpaper]{IEEEtran}
\usepackage{graphicx}
\usepackage{subfigure}
\usepackage{epstopdf}
\usepackage{comment}

\usepackage[cmex10]{amsmath}
\usepackage{amsfonts,amsthm}
\usepackage{algorithm}
\usepackage{algorithmic}
\usepackage{cite}
\usepackage{booktabs}

\theoremstyle{definition}

\newcommand{\ssf}[1]{\textrm{$\sf{#1}$}{}}
\usepackage{color}
\usepackage{bm}

\DeclareMathOperator*{\maximize}{maximize}
\DeclareMathOperator*{\subj}{subject\,to}

\DeclareMathOperator*{\argmax}{arg\,max}

\usepackage{url}

\ifodd 0
\newcommand{\congr}[1]{{\color{red}#1}}
\else
\newcommand{\congr}[1]{#1}
\fi

\ifodd 0
\newcommand{\congc}[1]{{\color{red}(Cong: #1)}}
\else
\newcommand{\congc}[1]{}
\fi

\begin{document}

\title
{Towards Optimal Power Control via Ensembling Deep Neural Networks}

\author{Fei~Liang,
	Cong~Shen,~\IEEEmembership{Senior Member,~IEEE,}
	Wei~Yu,~\IEEEmembership{Fellow,~IEEE,}
	and~Feng~Wu,~\IEEEmembership{Fellow,~IEEE}
	\thanks{F. Liang, C. Shen and F. Wu are with the Laboratory of Future Networks, School of Information Science and Technology, University of Science and Technology of China.}
	\thanks{W. Yu is with the Electrical and Computer Engineering Department, University of Toronto.}
}

\maketitle

\begin{abstract}
A deep neural network (DNN) based power control method is proposed, which aims at solving the non-convex optimization problem of maximizing the sum rate of a fading multi-user interference channel. Towards this end, we first present \emph{PCNet}, which is a multi-layer fully connected neural network that is specifically designed for the power control problem. A key challenge in training a DNN for the power control problem is the lack of ground truth, i.e., the optimal power allocation is unknown. To address this issue, PCNet leverages the unsupervised learning strategy and directly maximizes the sum rate in the training phase. We then present \emph{PCNet+}, which enhances the generalization capacity of PCNet by incorporating noise power as an input to the network. Observing that a single PCNet(+) does not universally outperform the existing solutions, we further propose \emph{ePCNet(+)}, a network ensemble with multiple PCNets(+) trained independently. Simulation results show that for the standard symmetric $K$-user Gaussian interference channel, the proposed methods can outperform all state-of-the-art power control solutions under a variety of system configurations. Furthermore, the performance improvement of ePCNet comes with a reduced computational complexity.

\end{abstract}

\IEEEpeerreviewmaketitle

\section{Introduction}

The capacity region of the multi-user interference channel is among the  longest outstanding open problems in information theory \cite{HanKobayashi,Tse2008}.  Progress on solving this problem has profound impact on today's wireless networks, as interference management has become the bottleneck of the overall system performance due to the broadcast nature of wireless communications and the dense deployment of base stations and mobiles, which create a heavily interfering environment. 

In this paper, we make progress on the power control problem that maximizes the sum rate of a multi-user interference channel, where each receiver is restricted to treating interference as noise (TIN). 
This problem (including some variations) is generally NP-hard, and has been investigated for decades. Due to its non-convex nature, {state-of-the-art} solutions in the literature often involve either exhaustive search (explicitly or implicitly) \cite{qian2009mapel,liu2012achieving} or iterative optimization of some approximate sub-problems \cite{chen2011round,chiang2007power,gjendemsjo2008binary,shi2011iteratively}. Performance, convergence, and complexity issues hinder the practicality of these solutions. In particular, how to achieve efficient power control when the number of users is {large} remains an open problem.

This work addresses power control from a different perspective. Instead of directly tackling the non-convex optimization problem {in an analytical fashion}, we leverage the recent advances in deep learning to investigate whether a data-driven method can achieve better performance with lower complexity. \congr{In particular, the proposed approach establishes a connection between the sum-rate maximization problem and minimizing a loss function in training a deep neural network (DNN), and relies on the efficient network training and ensembling mechanism to achieve near-optimal power control.}

Deep learning has had great success in computer vision, natural language processing and some other applications. Recent results also show that deep learning can be a promising tool in solving difficult communication problems, such as channel decoding \cite{nachmani2018deep, liang2018}, radio signal classification \cite{o2018over}, and channel estimation \cite{ye2017power}. For the considered power control problem, Sun \emph{et. al.} \cite{sun2017learning} recently proposed a neural network based method, in which the network is trained with the power control results of WMMSE \cite{shi2011iteratively} serving as the ground truth. This method has lower computational complexity compared to the original WMMSE, but the sum rate is also upper bounded by it. A natural question thus arises: in addition to the benefit of low complexity, can we also achieve better performance (in terms of larger sum rate) than the existing power control methods?

In this paper, we answer this question affirmatively by proposing a novel family of methods, called \emph{\underline{e}nsemble \underline{P}ower \underline{C}ontrol \underline{Net}work (ePCNet)}. There are two key ingredients in {ePCNet}. The first is that in order to simultaneously achieve higher sum rate and lower computation complexity than existing methods, we move away from the supervised learning method used in \cite{sun2017learning} and adopt an \emph{unsupervised learning} strategy. The resulting \emph{PCNet} is trained to directly maximize the system sum rate, as opposed to minimizing the loss against any sub-optimal method (such as WMMSE). \congr{This idea lifts the performance upper bound limitation of \cite{sun2017learning} and allows us to approach the ultimate ground truth, i.e., the globally optimal power control.} A variation to PCNet, called \emph{PCNet+}, further enhances its generalization capacity by allowing a single trained network to handle a range of noise power levels.

The second component is \emph{ensemble learning}. This is particularly useful as we observe that a single  {PCNet} may not \emph{universally} outperform all existing methods. The proposed {ePCNet}, however, creates multiple independent ``copies'' of {PCNet} and trains them separately, before forming an ensemble that selects the power profile which leads to the largest sum rate. We show via numerical simulations that combining DNN with ensemble learning results in a high-performance and low-complexity power control method that \congr{outperforms state-of-the-art methods in a variety of system configurations}. Furthermore, this method is especially efficient when the number of users is large, for which traditional methods do not always provide satisfactory results.

The rest of this paper is organized as follows. Related works are discussed in Section~\ref{sec:rel}. The system model is introduced in Section~\ref{sec:system_model}. In Section~\ref{sec:pcnet}, the main contributions of this paper, including the {PCNet(+)}  and {ePCNet(+)}, are explained in detail. Simulation results are given in Section~\ref{sec:simulation_srm} and \ref{sec:simlation_srm_qc}. The paper is concluded in Section~\ref{sec:conclusion}.

To honor the tradition in the machine learning community and support reproducible research, we have made our source code publicly available at: \url{https://github.com/ShenGroup/PCNet-ePCNet}.

\section{Related Works}
\label{sec:rel}
\subsection{Power control for interference management}
Interference management via power control is a critical technique towards efficient frequency reuse in both wireless networks (such as cellular systems \cite{gjendemsjo2008binary}) and wired networks (such as Digital Subscriber Lines \cite{hong2014signal}). This problem has several variations, including min-rate maximization \cite{foschini1993simple}, sum-rate or weighted sum-rate maximization (SRM or WSRM) \cite{luo2009duality,papandriopoulos2009scale}, QoS constrained power minimization \cite{yates1995integrated}, and some hybrid formulations \cite{chiang2007power}.  Different channel models are also considered, including the scalar interference channel (IC) model, the parallel IC model and the MIMO IC model \cite{hong2014signal}. However, regardless of the specific problem formulation, the resulting optimization problem is generally \emph{non-convex}, which makes the solution non-trivial and has sparked a lot of research.   In this section, we give a brief overview of related literature. A comprehensive survey can be found in \cite{chiang2008power,hong2014signal}.

The sum rate or weighted sum rate maximization problem is one of the most fundamental formulations of power control problems, and has been thoroughly investigated in the literature. For the scalar IC model, {Chen et al.} \cite{chen2011round} proposed a round-robin power control method to determine the transmit power user-by-user, by solving a polynomial system equation. Binary power control was carefully investigated in \cite{gjendemsjo2008binary}. The authors found that binary power control can provide near-optimal performance and a low-complexity greedy binary power control algorithm was proposed. Noting that binary power control can be equivalently viewed as a \emph{scheduling} problem, the authors in \cite{naderializadeh2014itlinq} studied user scheduling and proposed a distributed algorithm called \emph{ITLinQ}, which implicitly maximizes the sum rate by activating a subset of links in which TIN can approach the capacity region with a provable gap. 

For the parallel IC model where the spectrum is divided into several independent non-overlapping bands, the problem is more complicated. In \cite{luo2008dynamic,luo2009duality}, this problem was addressed via Lagrangian dualization. However, due to the non-convex nature of the primal problem, a duality gap exists. This gap vanishes when the number of channels goes to infinity \cite{luo2009duality}. Noting that if the power allocation of all other users is fixed, power control of the present user can be effectively solved via water-filling, the authors of \cite{yu2002distributed} proposed to execute water-filling iteratively until convergence. A modified iterative water-filling method was later proposed for resource allocation in the presence of crosstalk \cite{yu2007multiuser}. For MIMO IC channels, {Shi et al.} transformed the original problem into an equivalent weighted sum mean-square error minimization (WMMSE) problem \cite{shi2011iteratively}. The WMMSE problem is convex for each individual variable when others are fixed, which makes it easier to find a local optimal solution. The authors of \cite{cirik2015weighted} further adopted the WMMSE method to study the full-duplex MIMO interference channel. The WMMSE algorithm turns out to be related to a more powerful fractional programming approach that can be used for the joint power control, scheduling and beamforming optimization problem \cite{WeiYu1,WeiYu2,WeiYu3}.

The power control problem becomes more complex if additional QoS constraints are incorporated. In \cite{chiang2007power}, the authors proposed to maximize the weighted sum rate with explicit QoS constraints via geometric programming (GP). Under the high-SNR assumption, the original problem can be approximated as a standard GP. Without the high-SNR assumption, it can be transformed into a series of GP problems, where the SCALE algorithm was proposed in \cite{papandriopoulos2009scale} to solve a series of approximated convex problems.

\subsection{Applications of deep learning in communications} 
Inspired by its success in many areas, researchers have begun to study whether deep learning can help improve the design of communication algorithms \cite{o2017introduction,wang2017deep}. In recent years, many works in this field have appeared, targeting different aspects of physical-layer and MAC-layer designs. 

Channel decoding is the first topic that researchers have attempted to leverage deep learning for better error performance and lower complexity. The authors in \cite{nachmani2018deep} carried out a thorough investigation on deep learning decoding of linear codes. Based on the belief-propagation (BP) decoding architecture, they adopted the feed-forward and recurrent neural network models to build channel decoders by assigning different weights to the edges in the Tanner graph. With careful training, the neural network decoder can achieve better performance than the traditional BP decoder \cite{nachmani2018deep}. Since BP decoding has high complexity, neural network decoding has also been investigated for min-sum decoding \cite{lugosch2017neural}. Combined with the modified random redundant (mRRD) iterative algorithm \cite{dimnik2009improved}, further performance improvement can be obtained. For channel decoding under colored noise, Liang \emph{et. al.} proposed an iterative BP-CNN structure in which one convolutional neural network is concatenated with a standard BP decoder for noise estimation \cite{liang2018}. To handle long block length, \cite{cammerer2017scaling} proposed to divide the Polar coding graph into sub-blocks, and the decoder for each sub-codeword can be trained separately to reduce the overall complexity. 

Besides channel decoding, deep learning has the potential to improve the state of the art in other areas. Toward the end-to-end learning of communication systems, Dorner \textit{et al.}\cite{dorner2018deep} have demonstrated the feasibility to design a communication system whose transmitter and receiver are entirely implemented with neural networks. The authors of\cite{samuel2017deep} applied deep neural networks to MIMO detection, which results in comparable detection performance but with much lower complexity. To address the excessive feedback overhead, Wen \textit{et. al.} designed \emph{CsiNet,} which learns a transformation from CSI to a codeword and a corresponding inverse transformation \cite{wen2018deep}. In \cite{kim2018deep}, neural networks have been used to design an encoder and decoder for sparse coded multiple access (SCMA). The resulting encoder if \cite{kim2018deep} can automatically construct efficient codebooks. Deep learning is also considered for radio signal classification \cite{o2018over}, traffic type recognition \cite{o2016end}, channel estimation \cite{neumann2018learning}, and optical fiber communications \cite{karanov2018end}. 

Narrowing down to the power control problem, the authors in \cite{sun2017learning} have proposed a power control algorithm that is based on training a deep neural network with WMMSE as benchmarks. They showed that the new solution can approach WMMSE in terms of the sum rate performance, but with much lower complexity. However, with WMMSE serving as the ground truth, the performance of the neural network is also upper bounded by the achievable sum rate of WMMSE, which is not the global optimum. This limitation has motivated us to abandon the supervised learning strategy in training DNN, and resort to various new techniques, such as unsupervised learning, ensemble learning and batch normalization. The goal is to simultaneously improve the sum-rate performance over state-of-the-art algorithms while reducing the online computational complexity.

\section{System Model}
\label{sec:system_model}

\subsection{K-User single-antenna interference channel}
\begin{figure}
	\centering
	\includegraphics[width=0.3\linewidth]{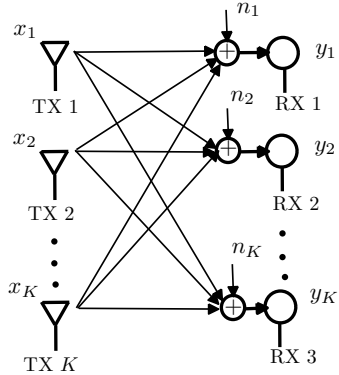}
	\caption{The $K$-user interference channel.}
	\label{fig:ic}
\end{figure}

We consider a general $K$-user single-antenna interference channel as shown in Fig. \ref{fig:ic}. It is assumed that all transmitter-receiver pairs share the same narrowband spectrum and are synchronized. The discrete-time baseband signal received by the $i$-th receiver is given by
\begin{equation}
y_{i} = h_{i,i}x_{i} + \sum_{j \in \mathcal{K}/\{i\}} h_{j,i}x_{j} + n_{i}, 
\label{eq1}
\end{equation}
where we let the set of transmitter-receiver pairs be $\mathcal{K} = \{1,2,\cdots,K\}$; $\mathcal{K}/\{i\}$ denotes the set of transmitter-receiver pairs excluding the $i$-th one; $x_{i} \in \mathbb{C}$ denotes the signal transmitted by the $i$-th transmitter; $h_{i,i} \in \mathbb{C}$ denotes the direct-link channel for the $i$-th user, $h_{j,i} \in \mathbb{C}$ denotes the cross-link channel between the $j$-th transmitter and the $i$-th receiver; and $n_{i} \sim \mathcal{CN}(0,\sigma_{i}^{2})$ denotes the receiver noise, which is independent across both time and users. Receiver $i$ only intends to decode $x_{i}$. For simplicity, we assume that all receivers have the same noise power $\sigma^2$. We note that this model has been widely studied in the literature; see \cite{shi2011iteratively,sun2017learning,qian2009mapel,chen2011round}.

A block fading channel model is assumed, i.e., the channel coefficients remain unchanged in one time slot but change independently from one time slot to another. We do not pose any limitation on the distribution of $h_{i,i}$, as our method is generic enough to handle different channel models.  Random Gaussian codebooks are assumed. Encoding is independent across users and no interference cancellation is performed at each receiver. Thus, the transmitter-receiver pairs do not cooperate and multiuser interference is treated as additive noise, i.e., TIN \cite{naderializadeh2014itlinq}. Therefore, the effective received noise power at the $i$-th receiver is $\sigma_i^2+\sum_{j \in \mathcal{K}/\{i\}}P_j\|h_{j,i}\|^2$.

The transmit power $P_{i}$ for user $i$ can be chosen from set $\mathcal{P}_{i} \subseteq \mathbb{R}_{+}$. In this work, for simplicity, we consider $\mathcal{P}_{i} = \{P: 0 \leq P \leq P_{\text{max}}\}, \forall i \in \mathcal{K}$, where $P_{\text{max}}$ is the maximum power that transmitters can use. Note that $0 \in \mathcal{P}_{i}$. Thus, a user may choose not to transmit\footnote{Hence, user scheduling is implicitly considered.}. The joint power profile of all users is denoted by ${\bf P} = \left(P_{1}, P_{2}, \cdots, P_{K}\right)^T \in  {\bf\mathcal{P}},$ where ${\bf \mathcal{P}} =  \mathcal{P}_{1} \times \mathcal{P}_{2} \times \cdots \times \mathcal{P}_{K}$ and $(\cdot)^T$ denotes matrix transpose.  For a given profile ${\bf P}$ and channel realizations $\left\{h_{ij}\right\}_{i,j \in \mathcal{K}}$, the achievable rate of the $i$-th receiver under Gaussian codebooks is
\begin{equation}
\label{eqn:user_rate}
R_i({\bf P})=\log\left(1+\frac{P_i\|h_{i,i}\|^2}{\sigma_i^2+\sum_{j \in \mathcal{K}/\{i\}}P_j\|h_{j,i}\|^2}\right).
\end{equation}
For each slot, the channel coefficients are fixed, and the power control algorithm outputs a power profile ${\bf P}$ based on the channel realization. 

\subsection{Problem formulation}
The objective of power control for interference management is to find the optimal power profile ${\bf P}$ for all users to maximize system performance under some specific constraints. With different performance measures and under different constraints, the power control problem has different formulations. In this paper, we focus on two specific power control problems: (1) sum rate maximization; and (2) sum rate maximization under QoS constraints. We comment that both formulations are widely researched in the literature; see \cite{chiang2008power} for a comprehensive survey.

\subsubsection{Sum Rate Maximization (SRM)}
The sum rate maximization (SRM) problem is formally given as
\begin{equation}
\label{eqn:rate_max_problem}
\begin{split}
\maximize_{\mathbf{P}} \quad &\sum_{i=1}^{K}R_i({\mathbf{P}}) \\
\subj \quad &0 \leq P_i \leq P_{\text{max}}, \forall i \in \mathcal{K}.
\end{split}
\end{equation}
Problem \eqref{eqn:rate_max_problem} is the simplest formulation among various power control problems. However, it is very difficult to solve due to its non-convex nature with respect to  the power profile. The optimization problem \eqref{eqn:rate_max_problem} is known to be NP-hard \cite{chiang2008power}.

\subsubsection{Sum Rate Maximization under QoS Constraints (SRM-QC)}
A more complicated variation of \eqref{eqn:rate_max_problem} is to maximize the sum rate while satisfying the minimum rate requirement of all receivers, which can be formally presented as
\begin{equation}
\label{eqn:rate_max_qos_constraint_problem}
\begin{split}
\maximize_{\mathbf{P}} \quad &\sum_{i=1}^{K}R_i({\mathbf{P}}) \\
\subj \quad &R_i({\mathbf{P}})\geq r_{i,\min}, \forall i \in \mathcal{K},\\
&0 \leq P_i \leq P_{\text{max}}, \forall i \in \mathcal{K},
\end{split}
\end{equation}
where $r_{i,\min}$ is the minimum required rate of the $i$-th receiver. For simplicity, we define $$\mathbf{r}_{\min}=(r_{1,\min}, \cdots, r_{K,\min}).$$
Obviously, if we set $r_{i,\min}=0,\forall i \in \mathcal{K}$, the SRM-QC problem \eqref{eqn:rate_max_qos_constraint_problem} degenerates to the SRM problem \eqref{eqn:rate_max_problem}.

If the target rates are large, the SRM-QC problem may have no feasible solution. It is not difficult to derive a criterion to check its feasibility \cite{chiang2008power,qian2009mapel}. Define matrix $\bf{B}$ as:
\begin{equation}\label{eqn:chk_mat}
\begin{split}
B_{i,j}=
\begin{cases}
0, & i=j\\
\frac{\gamma_{i,\min}\lVert h_{j,i}\rVert^2}{\lVert h_{i,i}\rVert^2}, & i\neq j,
\end{cases}
\end{split}
\end{equation}
where $B_{i,j}$ denotes the $(i,j)$-th element of $\bf{B}$,  and $\gamma_{i,\min}$ is the minimum  signal-to-interference-plus-noise ratio (SINR) of the $i$-th receiver that is required to satisfy the minimum rate constraint, i.e. $\gamma_{i,\min}=2^{r_i,\min}-1$. If the maximum eigenvalue of $\bf{B}$ is larger than 1, it is possible to find a feasible power allocation $\widehat{\bf{P}}$ as
\begin{equation}\label{eqn:feasible_sln}
\begin{split}
\widehat{\bf{P}} = \left(\bf{I-B}\right)^{-1}\bf{u},
\end{split}
\end{equation}
where $I$ denotes an $K\times K$ identity matrix and $\bf{u}$ is a $K\times 1$ column vector with the $i$-th element $u_i$ as
\begin{equation}
\begin{split}
u_i=\frac{\gamma_{i,\min}\sigma_i^2}{\lVert h_{i,i}\rVert}.
\end{split}
\end{equation}
If all elements in $\widehat{\bf{P}}$ are in the range between 0 and $P_{\max}$, then the power profile $\widehat{\bf{P}}$ is a feasible solution of the problem \eqref{eqn:rate_max_qos_constraint_problem}. Otherwise  \eqref{eqn:rate_max_qos_constraint_problem} is not feasible. Notes that $\widehat{\bf{P}}$ may not be the optimal solution.

\section{PCNet and PCNet+: Training DNN for Power Control}
\label{sec:pcnet}
We describe the proposed \emph{PCNet}, and a simple variation \emph{PCNet+}, in this section, including details of the DNN design and the training mechanism based on unsupervised learning. As mentioned before, the SRM problem \eqref{eqn:rate_max_problem} is a special format of the SRM-QC problem \eqref{eqn:rate_max_qos_constraint_problem}. We thus describe the proposed design that is universal to both problem \eqref{eqn:rate_max_problem} and \eqref{eqn:rate_max_qos_constraint_problem}, and only highlight the differences when applicable.

\subsection{Network design}
\begin{figure}
	\centering
	\includegraphics[width=0.85\linewidth]{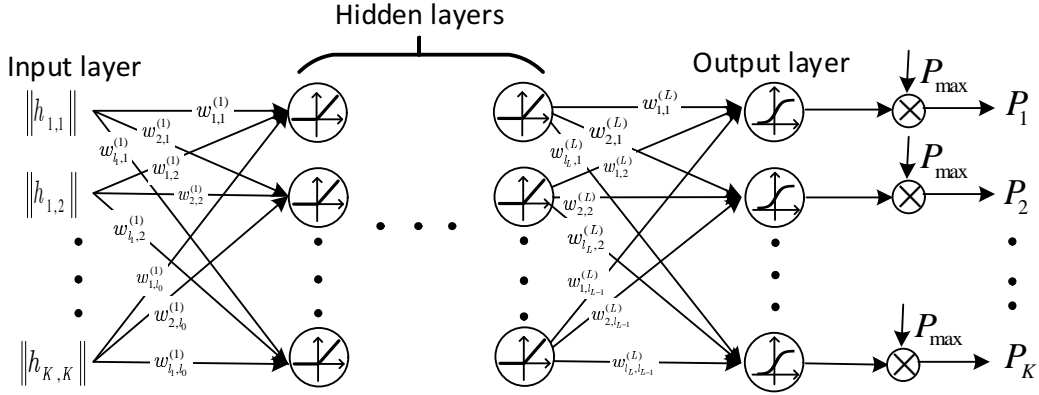}
	\caption{Illustration of \emph{PCNet}.}
	\label{fig:network}
\end{figure}
We propose to exploit a fully connected deep neural network to address the power control problem \eqref{eqn:rate_max_problem}. The network structure is illustrated in Fig.~\ref{fig:network}. More specifically, the network consists of one input layer of $K^2$ nodes, one output layer of $K$ nodes, and $L-1$ fully connected hidden layers. These layers are indexed from 0 to $L$. The input layer of PCNet  is formed by aligning all $\|h_{i,j}\|$ as a column vector, denoted as $\mathbf{h}$. The input vector is processed by the $L$ fully connected layers, including $L-1$ hidden layers and one output layer. 

The reason that a fully connected DNN is adopted, other than more sophisticated networks such as convolutional neural networks (CNN) or recurrent neural networks (RNN), is that there is little structure to explore for the general problem \eqref{eqn:rate_max_problem} and \eqref{eqn:rate_max_qos_constraint_problem}. On the other hand, if the problem exhibits certain features, such as the existence of correlation among different channel coefficients, then a more structured neural network such as CNN may be useful. This topic is left for consideration in a future work.

We denote the number of  nodes in the $k$-th layer as $l_k$. If the $k$-th layer is a hidden layer, its output is calculated as follows:
\begin{equation}\label{eqn:fc_layer}
\begin{split}
\mathbf{c}_k = \ssf{ReLU}\left(\ssf{BN}\left(\mathbf{W}_k\mathbf{c}_{k-1} + \mathbf{b}_k\right)\right),
\end{split}
\end{equation}
where $\mathbf{c}_{k-1}$ and $\mathbf{c}_{k}$ are the output vectors of the previous and current layers; their dimensions are $l_{k-1}\times1$ and $l_{k}\times1$ respectively; $\mathbf{W}_k$ is the $l_k\times l_{k-1}$ weight matrix and $\mathbf{b}_k$ is the $l_k\times 1$ bias vector; $\ssf{BN}(\cdot)$ denotes the batch normalization \cite{ioffe2015batch};  $\ssf{ReLU}(\cdot)$ is the Rectified Linear Unit function ($\max(x,0)$) which introduces nonlinearity to the network \cite{nair2010rectified}. For the first hidden layer, we define $\mathbf{c_0}=\mathbf{h}$ and $l_0=K^2$.

The output layer decides the transmit power of all transmitters. Here, the calculation is different from previous layers and is given as follows:
\begin{equation}\label{eqn:final_layer}
\begin{split}
\mathbf{c}_L = \ssf{Sig}\left(\mathbf{W}_L\mathbf{c}_{L-1} + \mathbf{b}_L\right),
\end{split}
\end{equation}
where $\ssf{Sig}(x)$ denotes the standard sigmoid function: $$\ssf{Sig}(x)=\frac{1}{1+\exp(-x)}.$$ Note that because the transmit power must be within the range $[0, P_{\text{max}}]$, the sigmoid function is used as the activation function instead of $\ssf{ReLU}$ to regulate the output. Batch normalization is not used in the output layer as we have empirically observed that it would degrade the network performance in our problem. Finally, the transmit power of user $i$ is obtained by
\begin{equation}\label{eqn:power_result}
\begin{split}
P_i = P_{\text{max}}c_{L,i},
\end{split}
\end{equation}
where $c_{L,i}$ is the $i$th element of $\mathbf{c}_L$.

We note that the fully connected DNN structure can be completely captured by the number of nodes in each layer and it is concisely denoted as:
\begin{equation}
\label{eqn:net_shape}
\begin{split}
\{l_0,l_1,l_2,...,l_L\}.
\end{split}
\end{equation}

\noindent\textbf{Remark:} A typical optimization problem involves one objective function and several constraints. PCNet essentially performs a mapping from a channel realization $\mathbf{h}$ to a power profile {\bf P}. However, it is not guaranteed that the output power profile is a feasible solution which satisfies all constraints, as these constraints may not always be explicitly enforced in the DNN structure. Thus measures have to be taken to address this problem. Fortunately, for the SRM problem formulated in \eqref{eqn:rate_max_problem}, the only constraint is that each transmit power must in the range $\left[0,P_{\max}\right]$. This simple constraint has naturally been taken care of in the proposed PCNet structure, by using $\ssf{Sig}(\cdot)$ as the activation function of the output layer, followed by a linear scaling of $P_{\max}$. The choice of $\ssf{Sig}(\cdot)$ and linear scaling directly regulate the PCNet output to satisfy the desired constraint. However, the minimum rate constraints in the SRM-QC problem are more difficult to handle. We are not aware of any neural network structure that can directly capture and enforce these constraints. We thus take a different approach and define a new loss function, which will be discussed in Section \ref{sec:trainPCnet}.

\subsection{Training PCNet}
\label{sec:trainPCnet}

The performance of a neural network largely depends on how it is trained. As mentioned before, in the recent work \cite{sun2017learning}, the authors proposed to train the network using WMMSE as the ground truth. The loss function is defined to measure the distance between the network output and the ground truth. Obviously, the network trained with this strategy cannot outperform WMMSE, and thus the main benefit of \cite{sun2017learning} comes from the (online) computational complexity. 

Ideally, were we able to obtain the globally optimal power control $\mathbf{P}^*=(P_1^*, \cdots, P_K^*)$ for a given channel realization, we would have designed PCNet under the traditional (supervised) neural network training by using the optimal solution as the ground truth and minimizing an appropriately chosen loss function (e.g., the L2 loss). However, such approach is suitable only when $P_i^*$ can be efficiently obtained, which is computationally challenging especially when $K$ is large.  This difficulty has motivated us to take a different approach, and to resort to \emph{unsupervised learning} in training PCNet. We directly apply the objective function in problems \eqref{eqn:rate_max_problem} and \eqref{eqn:rate_max_qos_constraint_problem} as the loss function for PCNet training, which is detailed in the following. 

\subsubsection{Training PCNet for SRM}
Noting that the sum rate, i.e. $\sum_{i=1}^{K}R_i(\bf{P})$, is the ultimate goal of the SRM problem, we can define a loss function to directly maximize the sum rate. More specifically, the loss function for training is defined as:
\begin{equation}
\label{eqn:loss_max_rate}
\ssf{loss}_{SRM} = -\mathbb{E}_{\mathbf{h}}\left[R(\mathbf{h}, \bm{\theta})\right],
\end{equation}
where $\bm{\theta}$ denotes the set of trainable network parameters for PCNet. $R(\mathbf{h}, \bm{\theta})$ is the sum rate under a specific channel realization $\mathbf{h}$ and network parameters $\bm{\theta}$. Obviously, this loss function is differentiable with respect to $\bm{\theta}$ and can be used to train the network via \congr{stochastic gradient descent (SGD)}.

Note that the loss function \eqref{eqn:loss_max_rate} is with respect to the distribution of $\mathbf{h}$, whose direct evaluation is quite challenging. We adopt a widely used method, \emph{mini-batch gradient descent}, to train PCNet under the loss function \eqref{eqn:loss_max_rate}. In each iteration of the training, multiple channel realizations $\mathbf{h}$ are generated from its distribution and we use $\mathcal{H}$ to denote the set of these channel realizations. The training loss is then approximated as
\begin{equation}\label{eqn:loss_approx}
\ssf{loss}_{\textit{SRM}}\approx -\frac{1}{\lvert\mathcal{H}\rvert}\sum_{\mathbf{h}\in \mathcal{H}}R(\mathbf{h},\bm{\theta}),
\end{equation} 
where $\lvert\mathcal{H}\rvert$ denotes the number of samples in $\mathcal{H}$ and its choice should balance complexity and accuracy. Therefore, in each iteration, \eqref{eqn:loss_approx} is used instead of \eqref{eqn:loss_max_rate} to calculate the gradients and update the network parameters.

\subsubsection{Training PCNet for SRM-QC}
For the SRM-QC problem, the loss function defined in $\eqref{eqn:loss_max_rate}$ is not applicable. Our proposed solution to address the minimum rate constraints in SRM-QC is to define a new loss function that penalizes the constraint violation. More specifically, the new loss function for SRM-QC is  defined as:
\begin{equation}
\label{eqn:loss_srm_qc}
\ssf{loss}_{\textit{SRM-QC}}
=\mathbb{E}_{\mathbf{h}}\left[-R(\mathbf{h}, \bm{\theta}) + \lambda\cdot\sum_{i=1}^{K}\ssf{ReLU}\left(r_{i,\min}-R_i(\mathbf{h}, \bm{\theta})\right)\right],
\end{equation}
where $R_i(\mathbf{h}, \bm{\theta})$ is the rate of $i$-th receiver under a specific channel realization $\mathbf{h}$ and network parameters $\bm{\theta}$.

Different from the loss function \eqref{eqn:loss_max_rate}, penalty terms are introduced to incentive the network output to meet the minimum rate constraints. If $r_{i,\min}>R_i(\mathbf{h}, \bm{\theta})$, i.e. a rate constraint is not satisfied, $\ssf{ReLU}\left(r_{i,\min}-R_i(\mathbf{h}, \bm{\theta})\right)>0$ and the corresponding penalty term will force the network parameters to be updated in the direction where the constraint is satisfied. On the contrary, if $r_{i,\min}\leq R_i(\mathbf{h}, \bm{\theta})$, $\ssf{ReLU}\left(r_{i,\min}-R_i(\mathbf{h}, \bm{\theta})\right)=0$ and the penalty term will not influence the network training. In this case, the training process will focus on making the network output satisfy the rate constraints of other receivers and increasing the system sum rate. The scaling factor $\lambda$ balances the trade-off of different terms in the loss function, which is a hyperparameter and needs to be tuned carefully: if it is too large, the network would focus on meeting the rate constraints while sacrificing the sum rate performance; if it is too small, the network may not output a feasible power profile. We will discuss how $\lambda$ influences the performance using numerical results in Section \ref{sec:simlation_srm_qc}.

It should be noted that even with the new loss function \eqref{eqn:loss_srm_qc}, PCNet may still output a power profile that is not a feasible solution. There are two possibilities when this happens. The first is that the original SRM-QC problem under this specific parameter setting is infeasible. In this case, there is no valid power profile anyway. The second is that feasible solutions exist but PCNet for SRM-QC fails to produce one. In this case, we note that the power profile given in \eqref{eqn:feasible_sln} is already a feasible solution. Therefore, we can obtain a feasible solution as
\begin{equation}\label{fig:normalized_feasible_solution}
\begin{split}
\widetilde{\mathbf{P}} = \widehat{\bf{P}} \cdot \frac{P_{\max}}{\max(\widehat{\bf{P}})},
\end{split}
\end{equation}
where $\max(\widehat{\bf{P}})$ denotes the maximum power allocated to one user in $\widehat{\bf{P}}$. Obviously, $\widetilde{\mathbf{P}}$ is also a feasible solution and there is at least one user whose power is $P_{\max}$. Previous result has shown that the achieved rate of $\widetilde{\mathbf{P}}$ is no smaller than $\widehat{\bf{P}}$ \cite{gjendemsjo2008binary}. If $\max(\widehat{\bf{P}})<P_{\max}$, the power profile $\widetilde{\mathbf{P}}$ would outperform $\widehat{\bf{P}}$.

\subsection{PCNet+: enhance the generalization capacity of PCNet}
\label{sec:pcnet+}
One PCNet needs to be trained for a given background noise power, because it only takes channel coefficients as the input.  In practical applications, this means that multiple network models of PCNet have to be separately trained for different  noise power (e.g., a range of SNRs), and the system controller needs to store multiple models and select one for power control based on its estimation of the noise power. This limits the generalization capability of PCNet. What is more desirable is a single trained PCNet which can handle a wide range of noise power levels.

Motivated by this limitation, we propose \emph{PCNet+}, which is a minor variation to PCNet but offers significantly better generalization capacity. The key idea of PCNet+ is to take the  noise power $\sigma^2$, in addition to the channel coefficients, as another input to the network. In this case, the network input layer contains $K^2+1$ nodes: $\|h_{i,j}\|,i=1,\cdots,K,j=1,\cdots,K$, and $\sigma^2$. In this way, a single PCNet+ can be trained to handle a range of noise power levels. In training PCNet+, the training data is generated equally under multiple levels of noise power, and the aggregated data is used. Simulation results, which are given in Section \ref{sec:simulation_srm},  will show that this method will enhance the generalization capacity of PCNet at the cost of very little performance loss.

\section{ePCNet: Ensembling PCNets}
\label{sec:epcnet}
Training PCNet with sufficiently representative data and the new loss functions defined in \eqref{eqn:loss_max_rate} (for SRM) or \eqref{eqn:loss_srm_qc} (for SRM-QC) cannot guarantee that PCNet outputs the globally optimal power profile $\mathbf{P}(\mathbf{h})^*$ for a given channel realization $\mathbf{h}$. Due to the inherent deficiency of gradient descent, the trained PCNet may fall into a local optimum. As will be corroborated in the numerical experiments, a single {PCNet} does not \emph{universally} outperform existing methods. 

\begin{figure}
	\centering
	\includegraphics[width=0.65\textwidth]{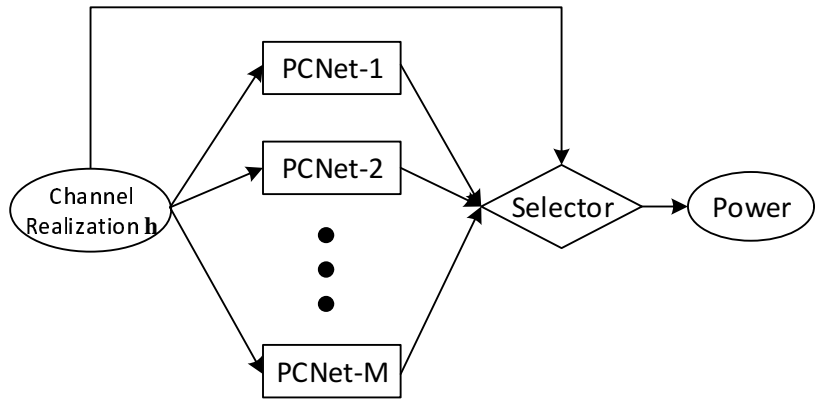}
	\caption{{ePCNet} with an ensemble of $M$ {PCNets}. }
	\label{fig:ensemble}
\end{figure}

To boost the power control performance and approach the global optimum, we  incorporate the idea of \emph{ensemble learning} and propose \emph{ensemble PCNet}, i.e., \emph{ePCNet}.  A pictorial illustration is given in Fig.~\ref{fig:ensemble}. Note that ensemble learning \cite{dietterich2002ensemble} is a commonly used machine learning method to achieve better performance by combining multiple \emph{weak} local learners, each of which is powerful only for certain local use cases. For example, in the classification problem, the ensemble output is obtained by weighing the intermediate outputs of all local classifiers in the ensemble. Similarly, for the power control problem, we propose to build an ensemble of PCNets to achieve a better performance. Although each individual PCNet is not ``powerful'' enough, i.e., cannot universally achieve better performance, a carefully formed ensemble of these weak local learners may be able to ``ride the peak'' of individual PCNets and output a universally better power profile.

Assume we have $M$ PCNets trained for a given interference channel model.  For simplicity, we assume that all PCNets have the same network structure. Operationally, this simplification  also facilitates the deployment of the proposed ePCNet, since all local learners share the same network structure and do not need to change the underlying computational architecture. We simply apply different network parameters to execute different PCNets. Each of the local PCNets is trained with a different set of initial parameters, together with a set of independently generated training data. These measures are taken to enhance the \emph{diversity} of the $M$ local learners, which is am important principle in ensemble learning \cite{dietterich2002ensemble}. The network input $\mathbf{h}$ is first fed into all PCNets in the ensemble and the $m$-th PCNet outputs a power control result $\mathbf{P}_m(\mathbf{h})$. The selector collects all $M$ outputs $\{ \mathbf{P}_m(\mathbf{h}), \forall m=1, \cdots, M \}$ as well as the channel vector $\mathbf{h}$, and then computes the sum rate for each PCNet. The selector outputs the power profile that corresponds to the highest sum rate, i.e.,
\begin{equation}
\begin{split}
\mathbf{P}_{ePCNet}(\mathbf{h})=\argmax_{\mathbf{P}_m(\mathbf{h})}R_{PCNet}(\mathbf{h},\mathbf{P}_m\left(\mathbf{h})\right),
\end{split}
\end{equation}
where $\mathbf{P}_{ePCNet}(\mathbf{h})$ denotes the output power profile of ePCNet, $R_{PCNet}(\mathbf{h},\mathbf{P}_m\left(\mathbf{h})\right)$ denotes the achieved sum rate of PCNet with the channel realization $\mathbf{h}$ and the power profile $\mathbf{P}_m(\mathbf{h})$.

There are three important remarks on the proposed ePCNet.  First, it is well-known that training a neural network with SGD may often lead to local optima. Thus, combining several local weak DNNs effectively looks at multiple local optima and leads to significantly improved sum rate performance.  Second, PCNet only requires a few matrix multiplications and is of low computational complexity\footnote{This is also observed and numerically validated in \cite{sun2017learning}.}. Therefore, combining several PCNets only increases the online complexity in a linear fashion with respect to the total number of local learners, which does not affect the order of complexity with respect to the problem size (i.e., $K$). Last but not the least, creating diversity by training multiple PCNets with different sets of parameters and training data is applicable not just to DNN. Some existing methods for power control may also benefit from such \emph{diversity}. For example, a random initialization is often needed to start an algorithm that \emph{iteratively} converges to a solution. Giving such algorithms multiple random initializations may also produce multiple local optima and selecting one with the best performance would also achieve better performance. However, such performance improvement is often achieved while sacrificing computational efficiency. In Section \ref{sec:simulation_srm}, we will compare the performance of ePCNet with that of WMMSE with multiple random initializations.

Finally, we remark that \emph{ePCNet+} can be easily obtained by ensembling several trained models of PCNet+, in the same fashion as constructing ePCNet from PCNets.

\section{Numerical Evaluations for SRM}
\label{sec:simulation_srm}
In the next two sections, we will conduct numerical simulations to verify the effectiveness of the proposed family of PCNets, and compare them with the state-of-the-art power control methods. More specifically, we intend to evaluate and compare the performance (for SRM and SRM-QC) and the computational complexity of various methods. Besides PCNet(+) and ePCNet(+), we also implement Round-Robin Power Control (RR), WMMSE and Greedy Binary Power Control (GBPC) for the SRM problem and the geometric programming (GP) based method \cite{chiang2007power} for the SRM-QC problem. In this section, we first focus on evaluations for SRM.

\subsection{Compared schemes}
\subsubsection{PCNet(+) and ePCNet(+)}

We implement the proposed PCNet and ePCNet in \emph{TensorFlow}, with the design following the detailed descriptions in Section~\ref{sec:pcnet} and \ref{sec:epcnet}. Following the common practice to train a neural network, we use Xavier initialization \cite{glorot2010understanding} to initialize the network parameters and ADAM \cite{kingma2014adam} to train the network. The stochastic gradient descent is used to calculate the gradient and the mini-batch size is set to $10^3$. To track the network performance during training, a validation data set containing $10^4$ samples is used to evaluate the network performance. This is done for every 50 iterations and the model with the best performance is saved. A total of $10^5$ iterations are executed to train the network.  The training data is randomly generated based on the channel model. As the mini-batch size is $10^3$ and total iteration number is $10^5$, a total of $10^8$ channel samples have been generated for training the network.

For SRM, it is observed that PCNet almost always outputs power profiles that are close to binary power control. Therefore, we round the power profiles of PCNet to binary ones. Binary power control facilitates practical implementation as it is easy for transmitters to configure binary transmit power. In addition, theoretical studies \cite{gjendemsjo2008binary} have shown that binary power control is optimal when $K=2$, and offers good performance when $K>2$ (although no longer optimal). Our result is consistent with these studies. However, for SRM-QC, we do not have these observations and the output of PCNet is used as the power profile directly.

\subsubsection{Round-Robin Power Control (RR)}
The RR algorithm proposed for SRM in \cite{chen2011round} is implemented for comparison. The basic idea of RR is to update the power of one user while keeping others fixed. This sub-problem is addressed by solving a polynomial function optimization. The algorithm stops when the following condition is satisfied,
\begin{equation}\label{eqn:stopping_criteria}
\frac{\lvert R(\mathbf{P}^{(t)}) - R(\mathbf{P}^{(t-1)})\rvert}{R(\mathbf{P}^{(t-1)})}\leq 10^{-4},
\end{equation}
where $R(\mathbf{P}^{(t)})$ and $R(\mathbf{P}^{(t-1)})$ denotes the sum rates in the current and last iterations respectively. 
\subsubsection{The Iteratively Weighted MMSE Approach (WMMSE)}
WMMSE is proposed to solve the sum-utility maximization problem for a MIMO interfering broadcast channel \cite{shi2011iteratively}. Its simplified version can be used to solve the considered SRM problem. In the literature, WMMSE is also used to generate the ground truth for the network training in \cite{sun2017learning}. Therefore, we also incorporate WMMSE for comparison. Note that the same stopping condition \eqref{eqn:stopping_criteria} is also used for WMMSE.
\subsubsection{Greedy Binary Power Control (GBPC)}
The greedy binary power control method in \cite{gjendemsjo2008binary} is also implemented for comparison. It has been shown in \cite{gjendemsjo2008binary} that GBPC can provide desirable performance that approaches the optimal binary power control.

Binary power control is actually a scheduling algorithm. There still exist some works addressing interference management via scheduling, such as \cite{naderializadeh2014itlinq,yi2016optimality}. The method proposed in \cite{nachmani2016learning} tries to activate a subset of links in which TIN can achieve the whole capacity region to within a constant gap. We empirically find it does not perform well in our simulations and the results are omitted in this paper. The work in \cite{yi2016optimality} maximizes the weighted sum generalized degrees of freedom, which is different from our target. Therefore we do not consider this work in comparison either.

\subsection{Performance comparison}
In majority of the simulations that follow, we focus on the symmetric interference channel model with i.i.d. Rayleigh fading for all channels. We also report performance comparison with Ricean fading and large-scale pathloss in Section~\ref{sec:sim_rician}.  For the symmetric Rayleigh fading IC model, we have $h_{i,j}\sim \mathcal{CN}(0, 1), \forall i, j \in \mathcal{K}$. The noise power at each receiver is set to the same $\sigma^2$. {These settings are widely used in the literature \cite{shi2011iteratively,sun2017learning,qian2009mapel,chen2011round}.} Without loss of generality, we assume $P_{\text{max}}=1$ and define the signal-to-noise ratio (SNR)  as
\begin{equation}
\label{eqn:def_EsN0}
\begin{split}
\ssf{EsN0} = 10\log\left(\frac{P_{\text{max}}}{\sigma^2}\right).
\end{split}
\end{equation}
{For PCNet/ePCNet, we present the results under three typical values of $\ssf{EsN0}$, 0dB, 5dB and 10dB. For PCNet+/ePCNet+, the network is trained using the data generated with $\ssf{EsN0}$ selected from the set $\{0,1,2,\cdots,10\}$dB. In each epoch of training, the numbers of training samples under different $\ssf{EsN0}$ are the same. We then test the performance of PCNet+/ePCNet+ with $\ssf{EsN0}$ as 0dB, 5dB and 10dB and the results are presented for comparison with PCNet/ePCNet trained for one speicific $\ssf{EsN0}$.}

We first focus on evaluating ePCNet for the SRM problem. In this case, the considered compared algorithms include RR, WMMSE and GBPC. We use $R_{PN}(\mathbf{h})$ to denote the achievable sum rate of one PCNet or ePCNet, given channel coefficients $\mathbf{h}$. Correspondingly, $R_{c}(\mathbf{h})$ denotes the sum-rate of a compared scheme. We evaluate the performance of ePCNet from different perspectives.

\begin{figure*}
	\centering
	\subfigure[$K=10, \ssf{EsN0}=0$dB]{\includegraphics[width=0.32\textwidth]{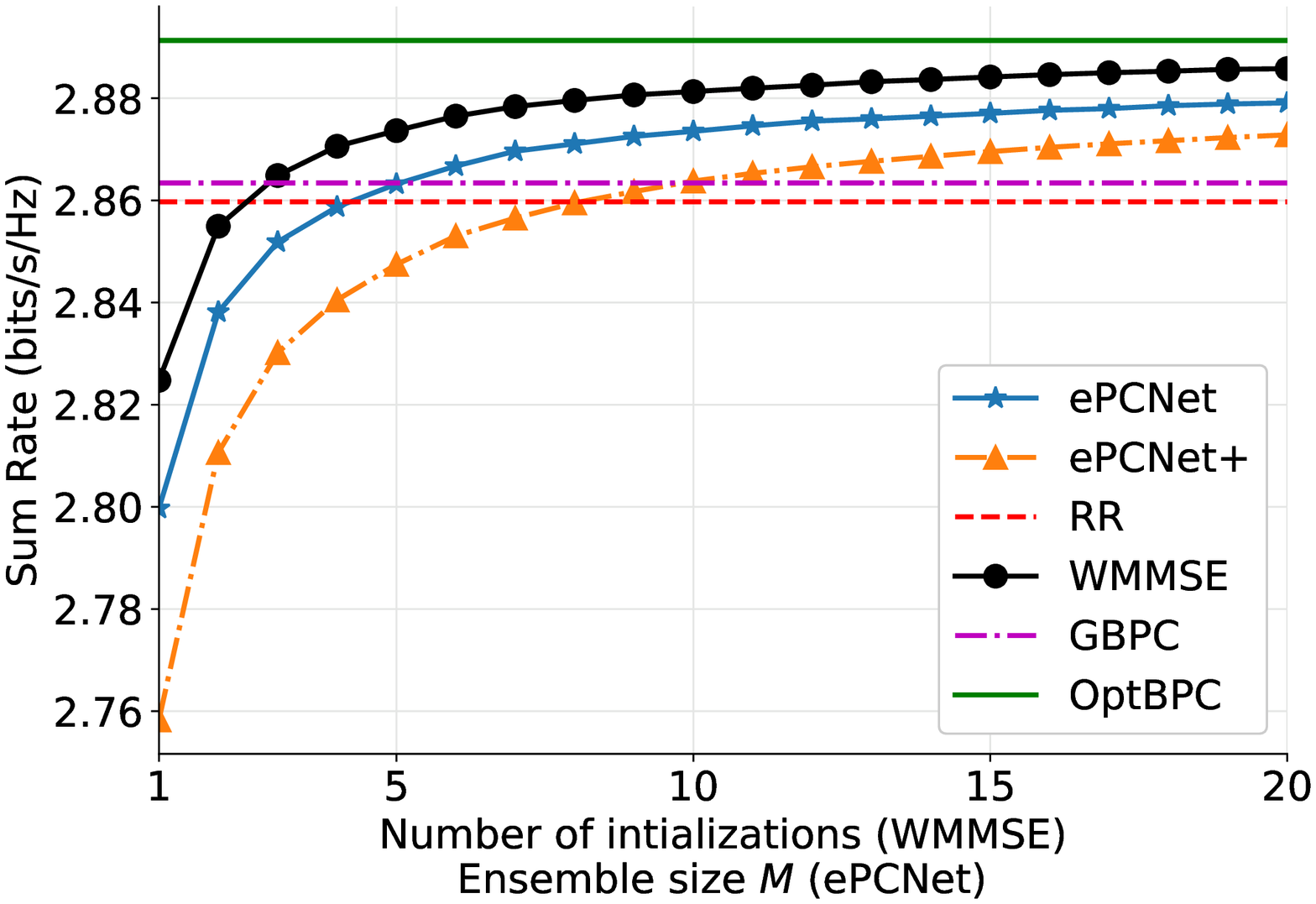}\label{fig:perf_comp_10_0}}
	\subfigure[$K=10, \ssf{EsN0}=5$dB]{\includegraphics[width=0.32\textwidth]{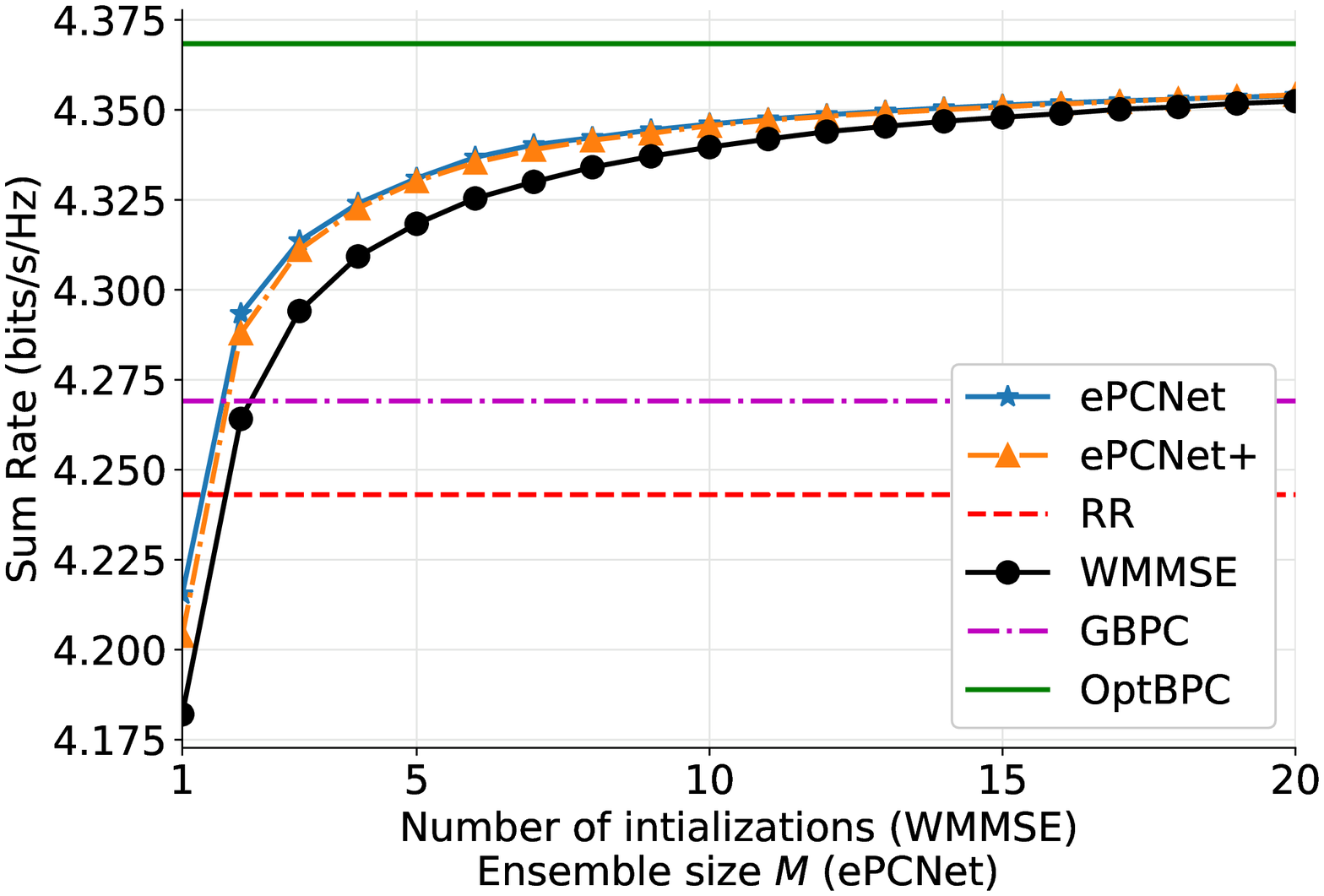}\label{fig:perf_comp_10_5}}
	\subfigure[$K=10, \ssf{EsN0}=10$dB]{\includegraphics[width=0.32\textwidth]{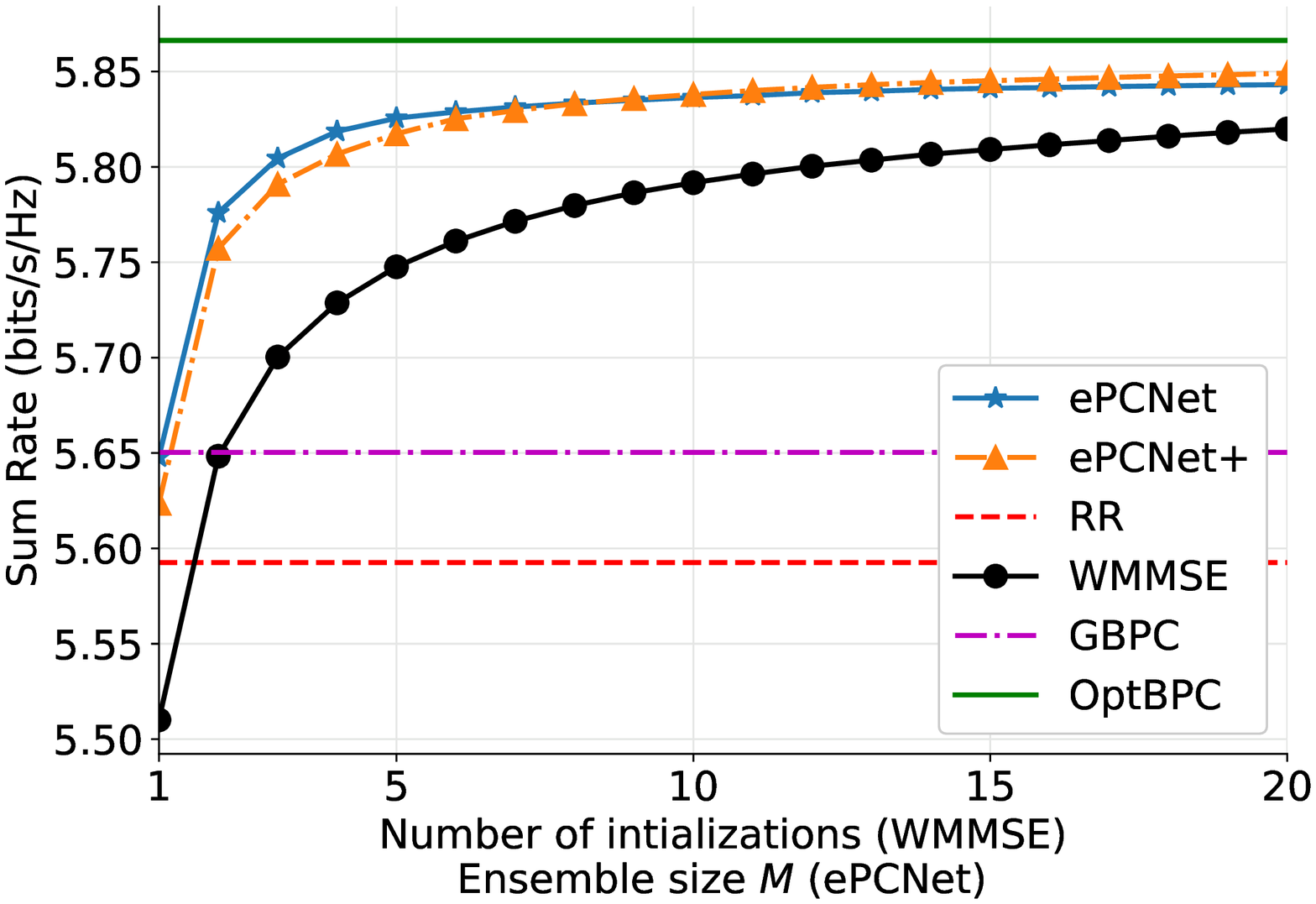}\label{fig:perf_comp_10_10}}
	\subfigure[$K=20, \ssf{EsN0}=0$dB]{\includegraphics[width=0.32\textwidth]{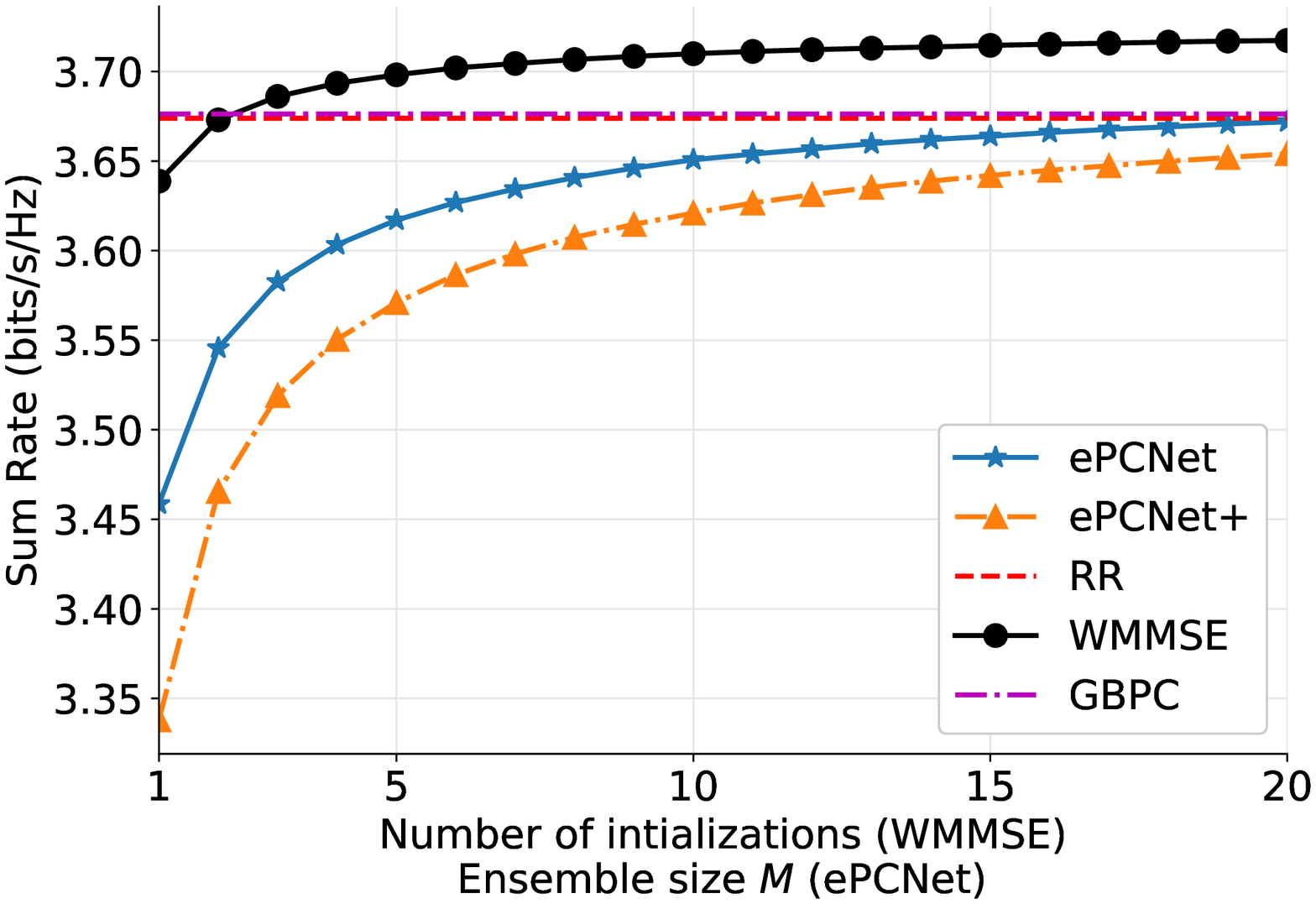}\label{fig:perf_comp_20_0}}
	\subfigure[$K=20, \ssf{EsN0}=5$dB]{\includegraphics[width=0.32\textwidth]{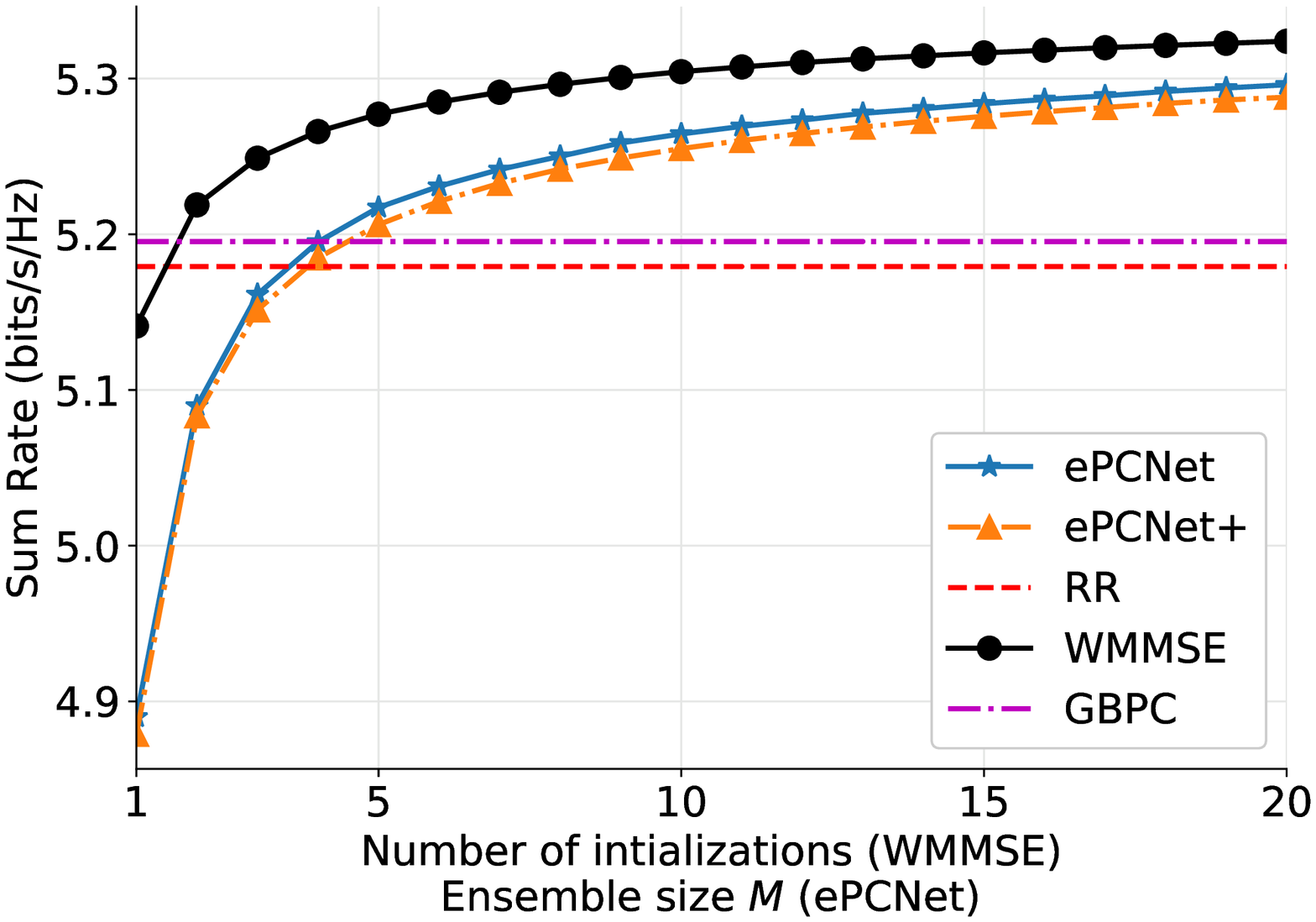}\label{fig:perf_comp_20_5}}
	\subfigure[$K=20, \ssf{EsN0}=10$dB]{\includegraphics[width=0.32\textwidth]{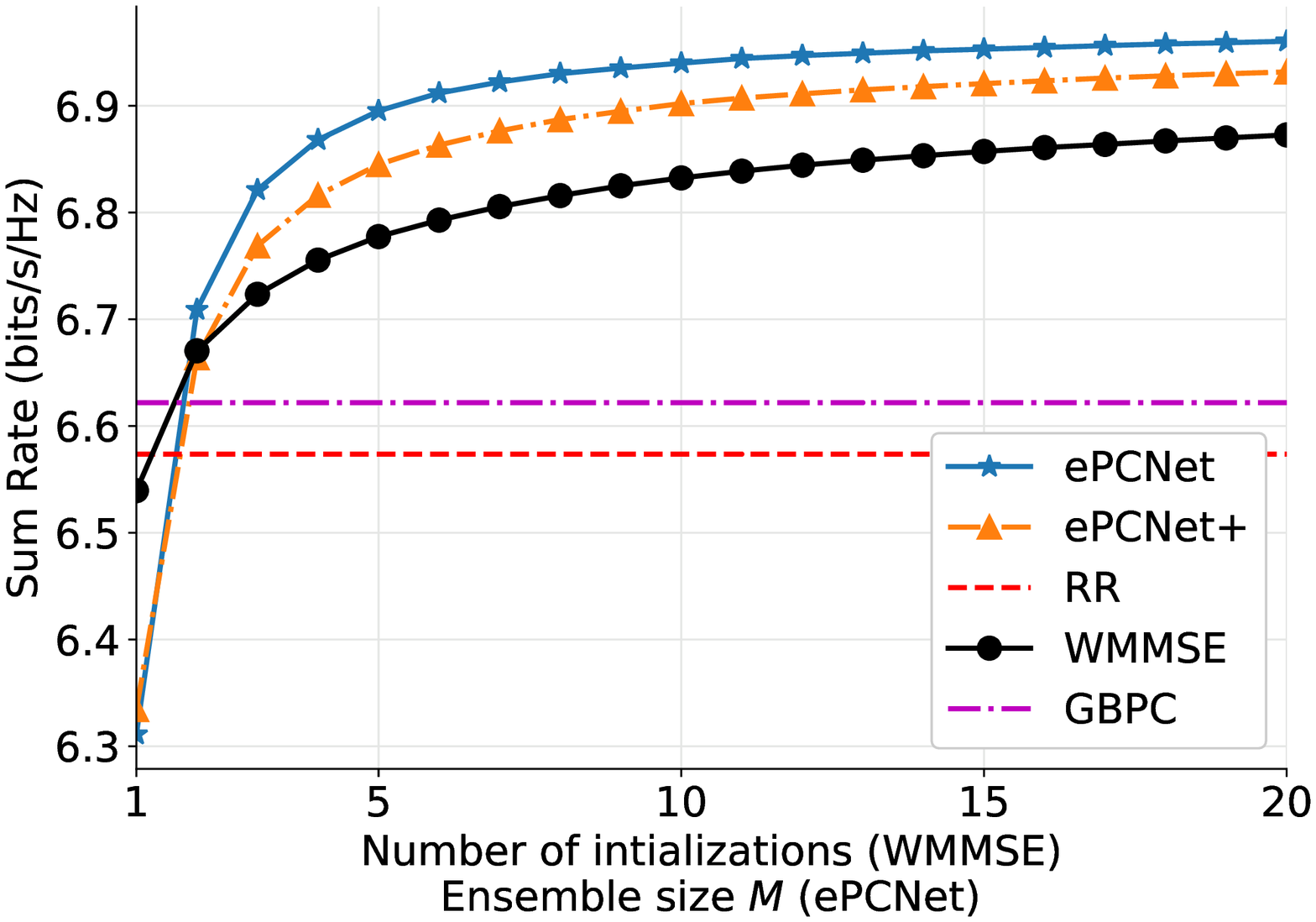}\label{fig:perf_comp_20_10}}
	\caption{Sum rate comparison of different power control methods. Left column is for $K=10$ and right column is for $K=20$. The legend \emph{OptBPC} denotes the optimal binary power control via an exhaustive search. }
	\label{fig:perf_comp}
\end{figure*}

First, we focus on comparing the achievable average sum rates of all methods, i.e., $\mathbb{E}_{\mathbf{h}}(R_{PN}(\mathbf{h}))$ and $\mathbb{E}_{\mathbf{h}}(R_{c}(\mathbf{h}))$. The results are shown in Fig.~\ref{fig:perf_comp} under different system settings. Specifically, for evaluating ePCNet, we show the performance of a network ensemble of different sizes to highlight how the overall performance of ePCNet (the ensemble)  scales with the number of PCNets (its local learners). The network structures for PCNet are $\{400,400,200,20\}$ and $\{100,200,100,10\}$ for $K=20$ and $K=10$, respectively. 

As we mention before, WMMSE is an iterative method and it needs a random initialization to start the algorithm. If complexity is not an issue, giving WMMSE multiple random initializations would also generate multiple local optima and selecting one with the best performance would also be a strategy to achieve better performance. Therefore, as comparison, we also evaluate the performance of WMMSE with multiple random initializations.

RR is an iterative method for which multiple initializations is also possible. However, as we will see later, the RR algorithm finds a local optimum by solving a high-order equation of $2K-2$ degrees and is of high computational complexity. Thus, we do not consider multiple initializations for the RR algorithm since it is not easy to do the experiments. As for GBPC, it is a deterministic algorithm and no randomness is contained. Hence, multiple random initializations are not possible for GBPC. The performances of RR and GBPC are plotted as horizontal lines for comparison.  When $K$ is not very large, it is possible to obtain the optimal binary power control via exhaustive search\footnote{There exist some methods in the literature to reduce the complexity of searching for the optimal binary power control \cite{gjendemsjo2008binary,liu2012achieving}.  However, they cannot fundamentally change the complexity scaling and hence are still limited to small $K$.}. We thus also plot the performance of optimal binary power control for $K=10$ in Fig. \ref{fig:perf_comp}.  

{Let us first compare the proposed PCNet/ePCNet and PCNet+/ePCNet+. It is clear that very little performance loss is observed for PCNet+/ePCNet+, compared with PCNet/ePCNet which is trained for the matching $\ssf{EsN0}$. In some cases such as Fig. \ref{fig:perf_comp_10_5}, \ref{fig:perf_comp_10_10} and \ref{fig:perf_comp_20_5}, two schemes provide almost identical results. This shows that PCNet+/ePCNet+ enjoys high generalization capacity while providing good adaption to different $\ssf{EsN0}$.}

Then we focus on comparing PCNet/ePCNet and WMMSE with multiple random initializations. A general observation we can get is that PCNet/ePCNet can outperform WMMSE when the user number is not too large and when SNR is high. As shown in Fig. \ref{fig:perf_comp_10_5}, Fig. \ref{fig:perf_comp_10_10} and Fig. \ref{fig:perf_comp_20_10}, the ePCNet with ensemble size as 10  outperforms WMMSE with 10 random initializations by 0.14\%, 0.77\% and 1.57\% respectively in these three cases. We comment that the performance improvement of WMMSE with multiple random initializations comes with significantly larger (online) complexity. If the computational resource is constrained and only one initialization is allowed, the performance gain of ePCNet becomes more obvious. As shown in Fig. \ref{fig:perf_comp_10_10} and Fig. \ref{fig:perf_comp_20_10}, ePCNet with 10 networks can outperform WMMSE by 5.92\% and 6.12\% in these two cases.

Due to the inherent difficulty in analyzing neural networks, we cannot give an accurate explanation to the above observations. Some intuitive explanations can be made. First, ePCNet does not necessarily achieve a performance gain when the user number $K$ is large, as shown in Fig. \ref{fig:perf_comp_10_5} and Fig. \ref{fig:perf_comp_20_5}. This can be explained by the fact that training a neural network of large size requires much more data, so it may not be an easy task. Second, WMMSE performs quite well when SNR is low. One possible explanation is that the background noise can smooth over those local optima of the optimization space. Thus at low SNR WMMSE does not suffer severely from being stuck at a local optimum. {However, it has been observed and reported in the deep learning literature that for multilayer networks such as DNN, different local optima often provide similar performance, even with highly non-convex loss functions. The authors of \cite{choromanska2015loss} have explained this phenomenon under several assumptions. Therefore, DNN may not be stuck at a bad local optimum and can outperform WMMSE in this case.}

\begin{figure*}
	\centering
	\subfigure[$K=10, \ssf{EsN0}=10$dB\label{fig:cdf_rate_10_10}]{\includegraphics[width=0.32\textwidth]{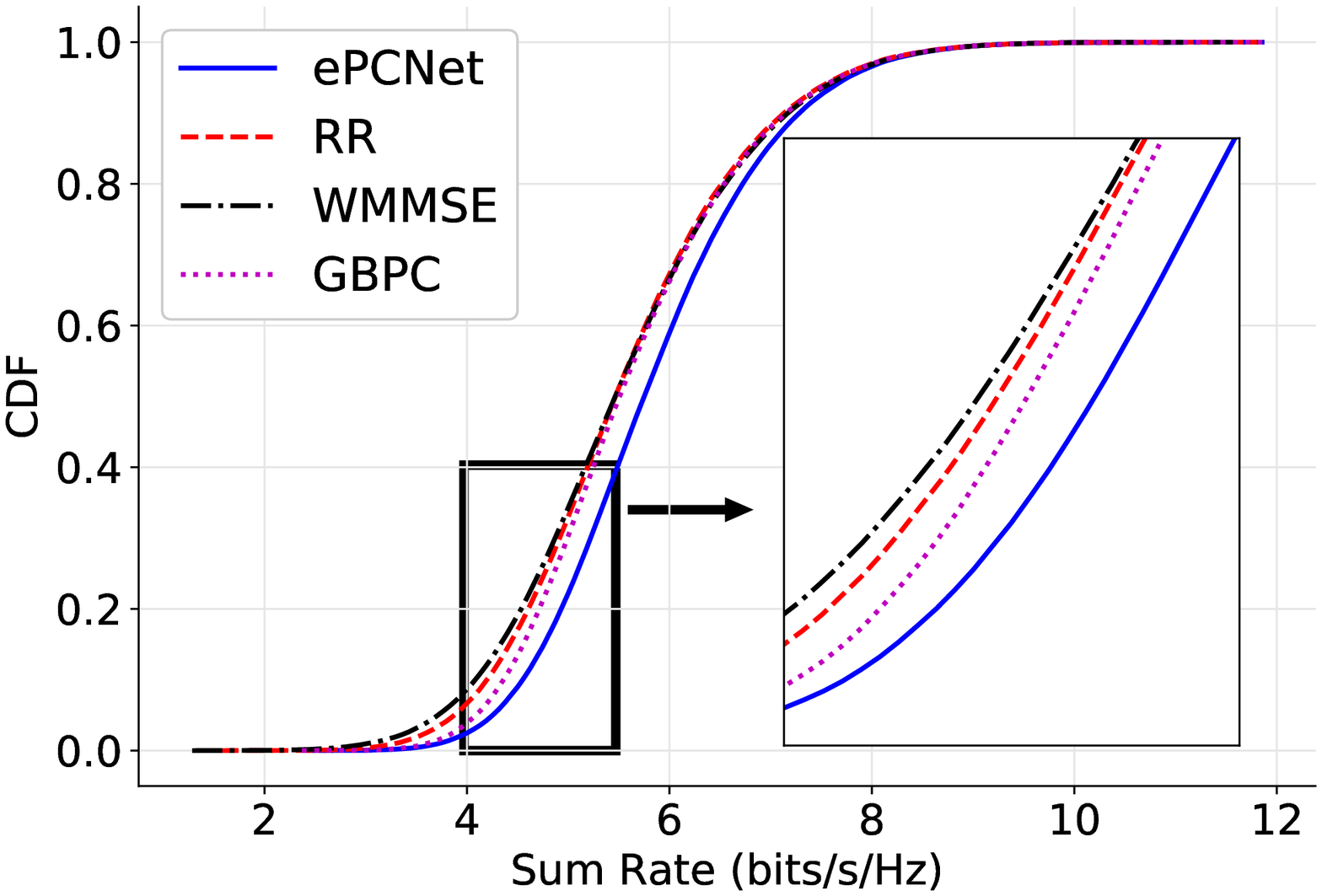}}
	\subfigure[$K=20, \ssf{EsN0}=10$dB\label{fig:cdf_rate_20_10}]{\includegraphics[width=0.32\textwidth]{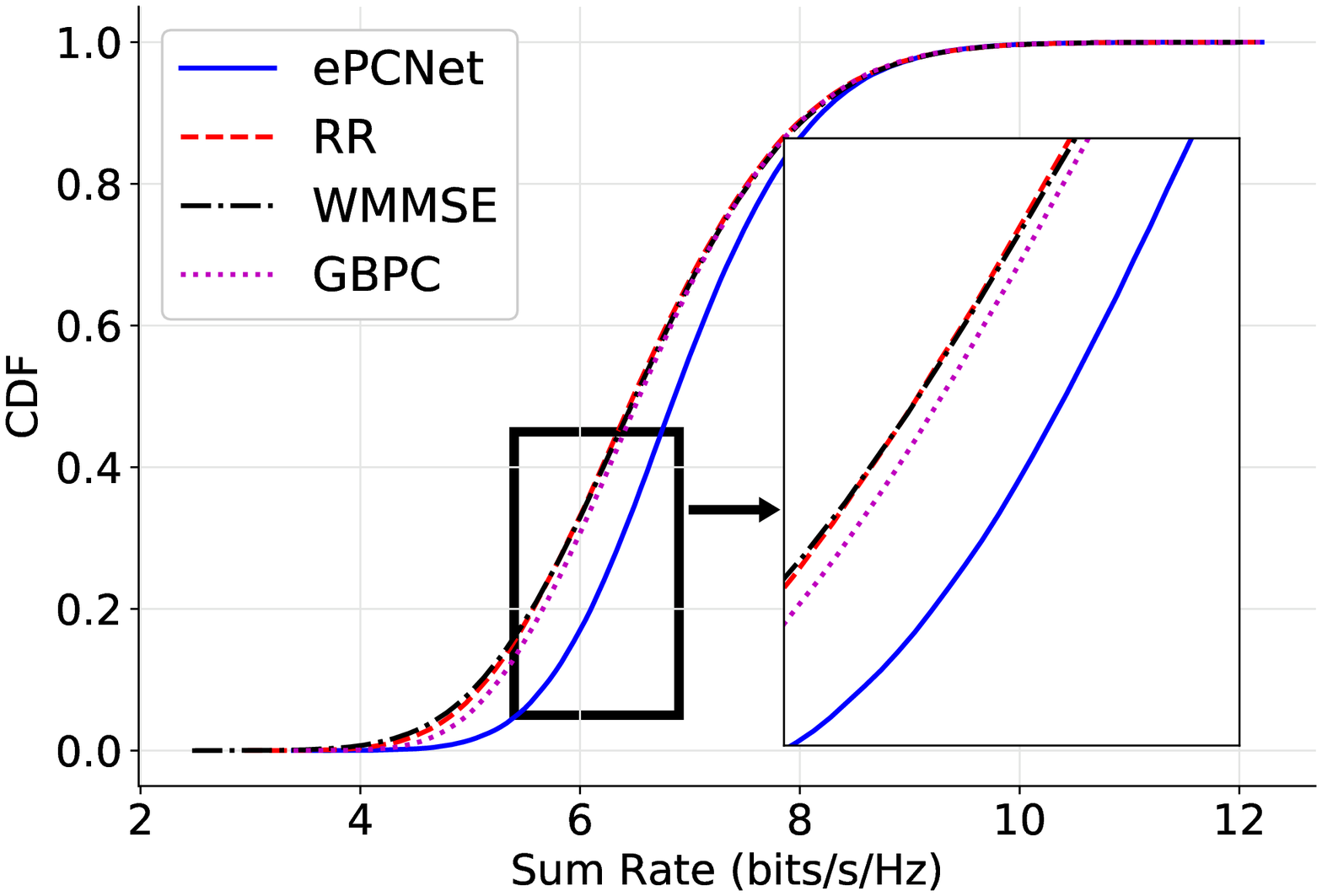}}
	\subfigure[$K=20, \ssf{EsN0}=5$dB\label{fig:cdf_rate_20_5}]{\includegraphics[width=0.32\textwidth]{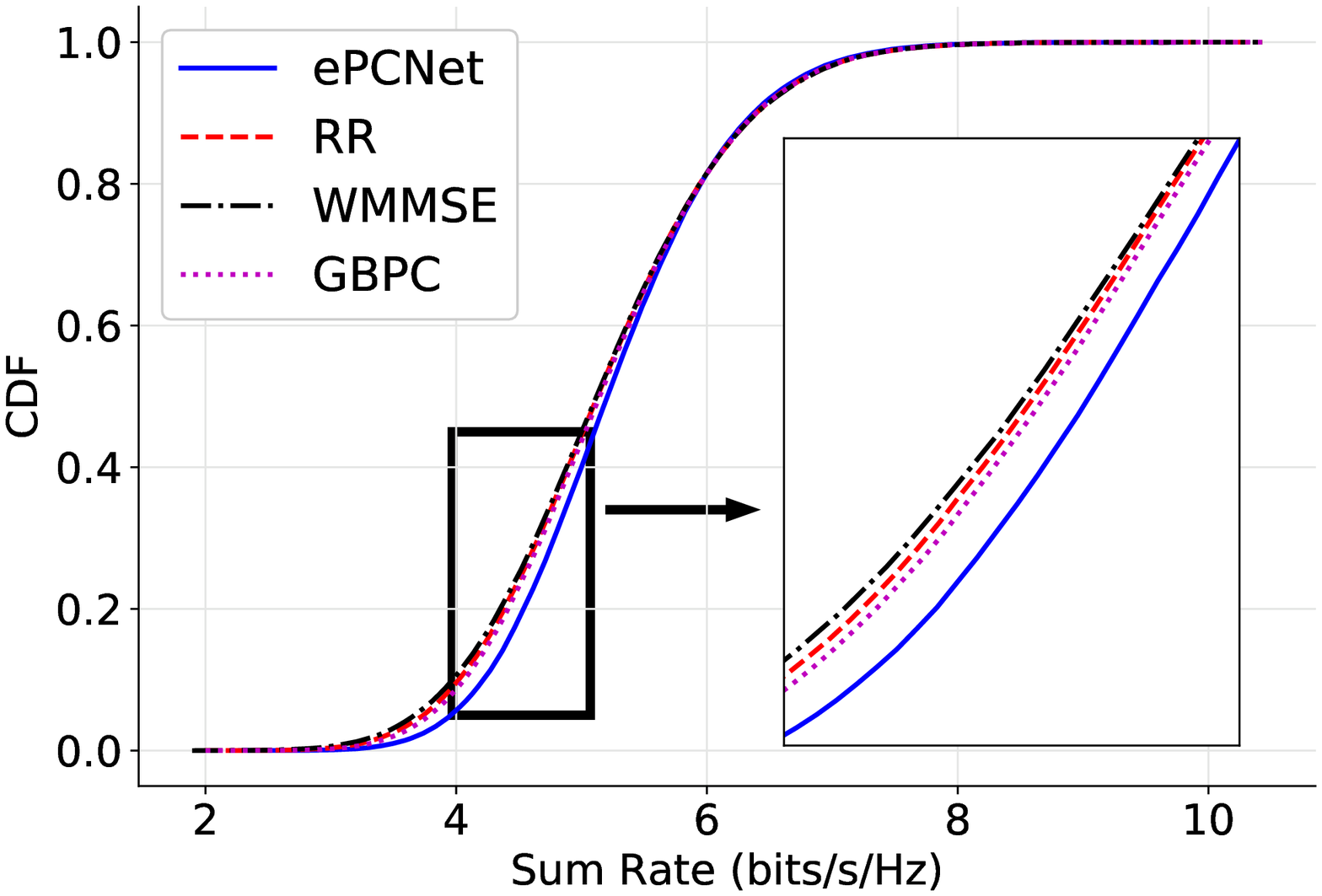}}
	\caption{Empirical CDF of achievable sum rates of different schemes.}
	\label{fig:cdf_rate}
\end{figure*}

Let us continue looking into the performance comparison between PCNet/ePCNet and other algorithms. It can be observed from Fig.~\ref{fig:perf_comp} that GBPC achieves the best performance among the three traditional methods if only one local optima is allowed in WMMSE and RR. When further compared to PCNet and ePCNet, we have the following observations. First, a single PCNet cannot achieve universally better sum rate than existing methods. There exists a performance gap between GBPC and a single PCNet. Second, when ePCNet is used and the ensemble size increases, it can quickly outperform standard WMMSE (with single initiation), RR, GBPC in most cases except the case $K=20,\ssf{EsN0}=0$dB. When $K=20, \ssf{EsN0}=10$dB, ePCNet with $M=2$ is already capable of outperforming GBPC. Even in the relatively difficult case of $K=20, \ssf{EsN0}=5$dB, ePCNet with $M=5$ has the best performance. Third, taking the three system settings shown in Fig.~\ref{fig:perf_comp_10_10}, Fig.~\ref{fig:perf_comp_20_10}, and Fig.~\ref{fig:perf_comp_20_5} as examples, the sum rate gain of ePCNet with $M=10$ over GBPC is 3.5\%, 4.6\% and 1.2\%, respectively. Further increasing $M$ continues to improve the performance, but the sum rate gain becomes marginal. Furthermore, by examining Fig.~\ref{fig:perf_comp_10_10}, we can see that ePCNet achieves near-optimal performance with respect to the optimal binary power control. Last but not the least, to take a closer look and compare the performance of all schemes, we also plot the empirical cumulative distribution functions (CDF) of achievable sum rates of all algorithms in Fig. \ref{fig:cdf_rate} under three settings. For ePCNet, we only show the performance of the ensemble with $M=10$ networks. For WMMSE, we show its performance without multiple random initializations. Clearly, the same conclusion can be reached that ePCNet outperforms the three baselines.

\begin{figure*}
	\centering
	\subfigure[ePCNet v.s. RR\label{fig:cdf_comp_rr}]{\includegraphics[width=0.32\textwidth]{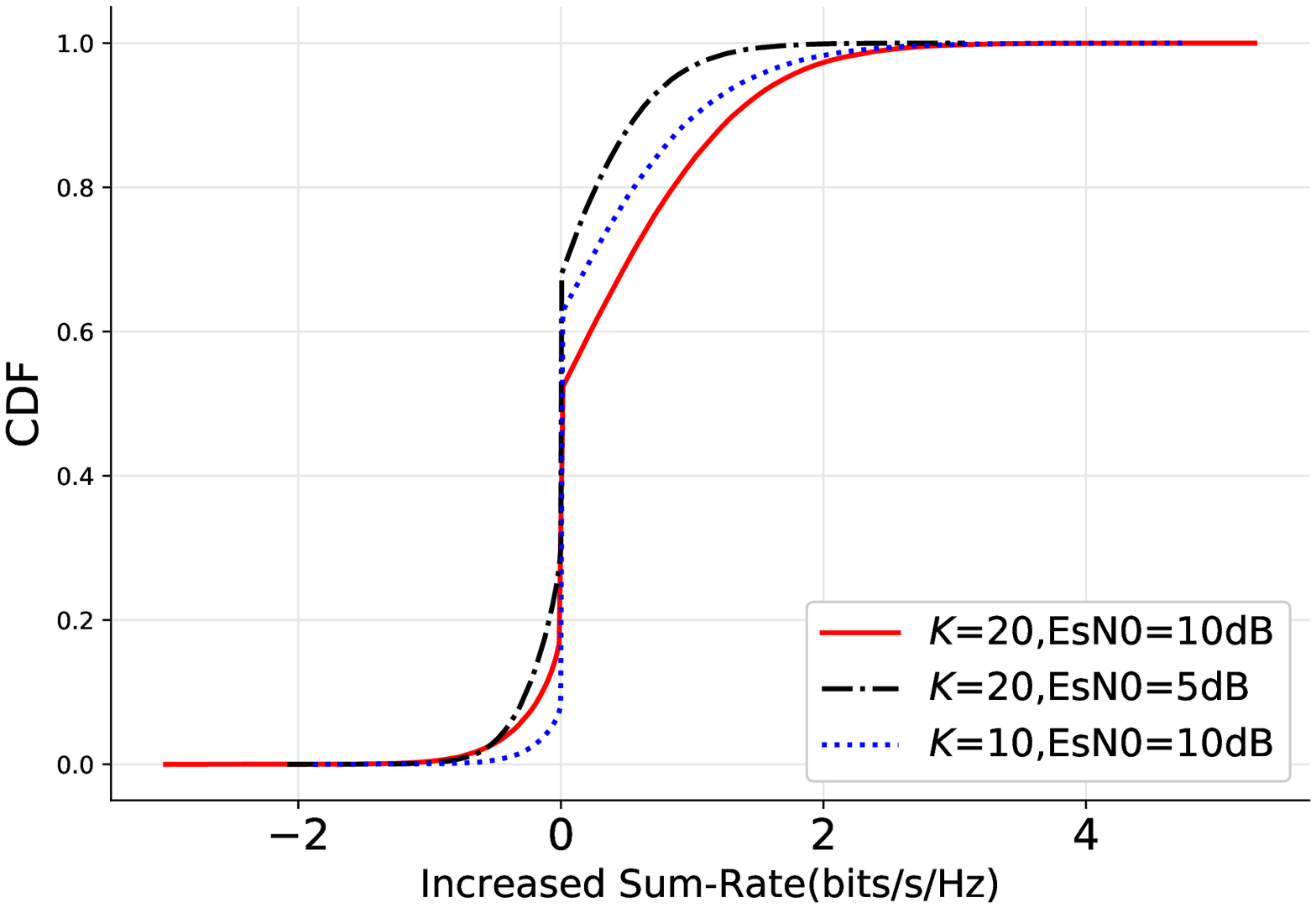}}
	\subfigure[ePCNet v.s. WMMSE\label{fig:cdf_comp_wmmse}]{\includegraphics[width=0.32\textwidth]{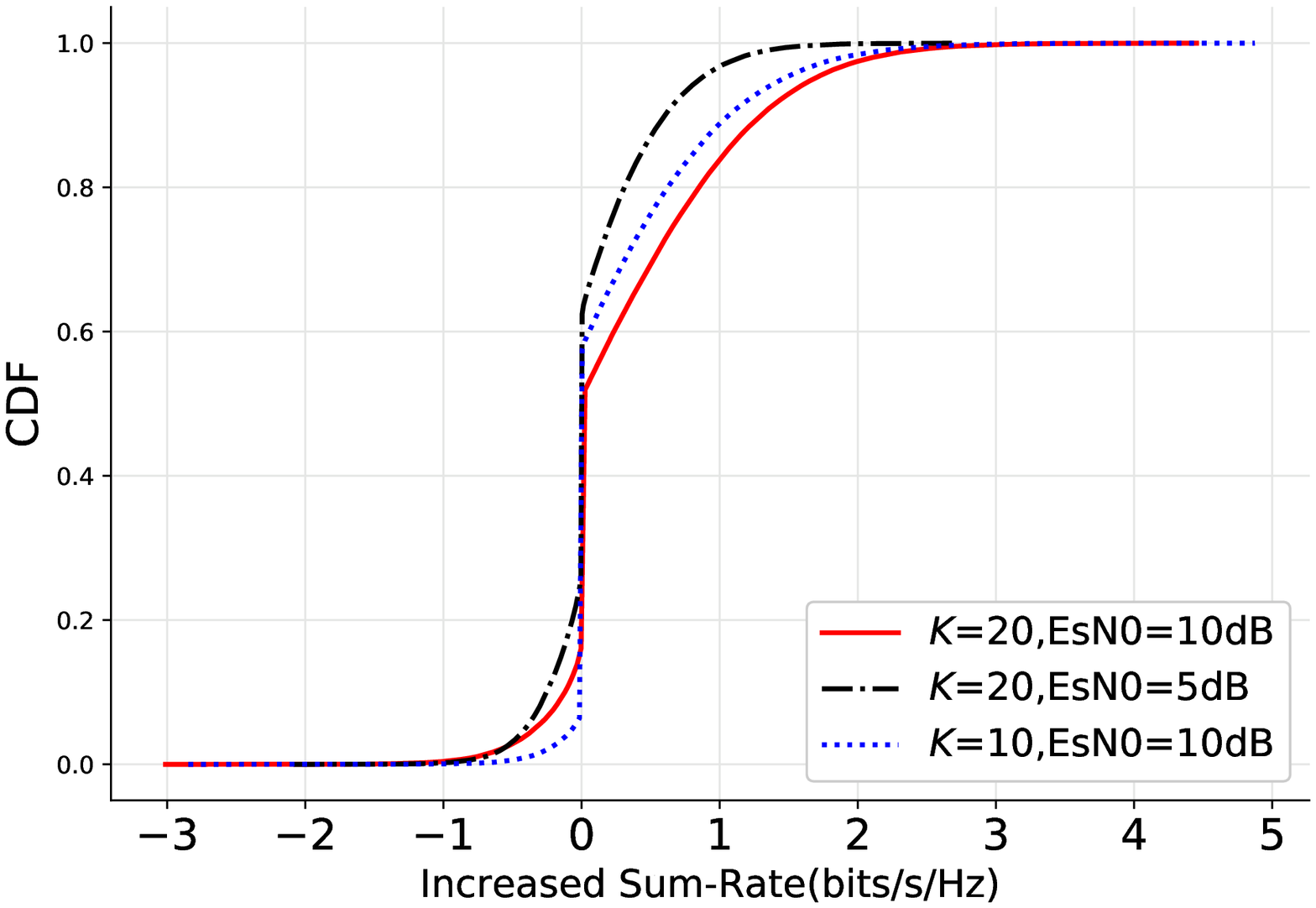}}
	\subfigure[ePCNet v.s. GBPC\label{fig:cdf_comp_gbpc}]{\includegraphics[width=0.32\textwidth]{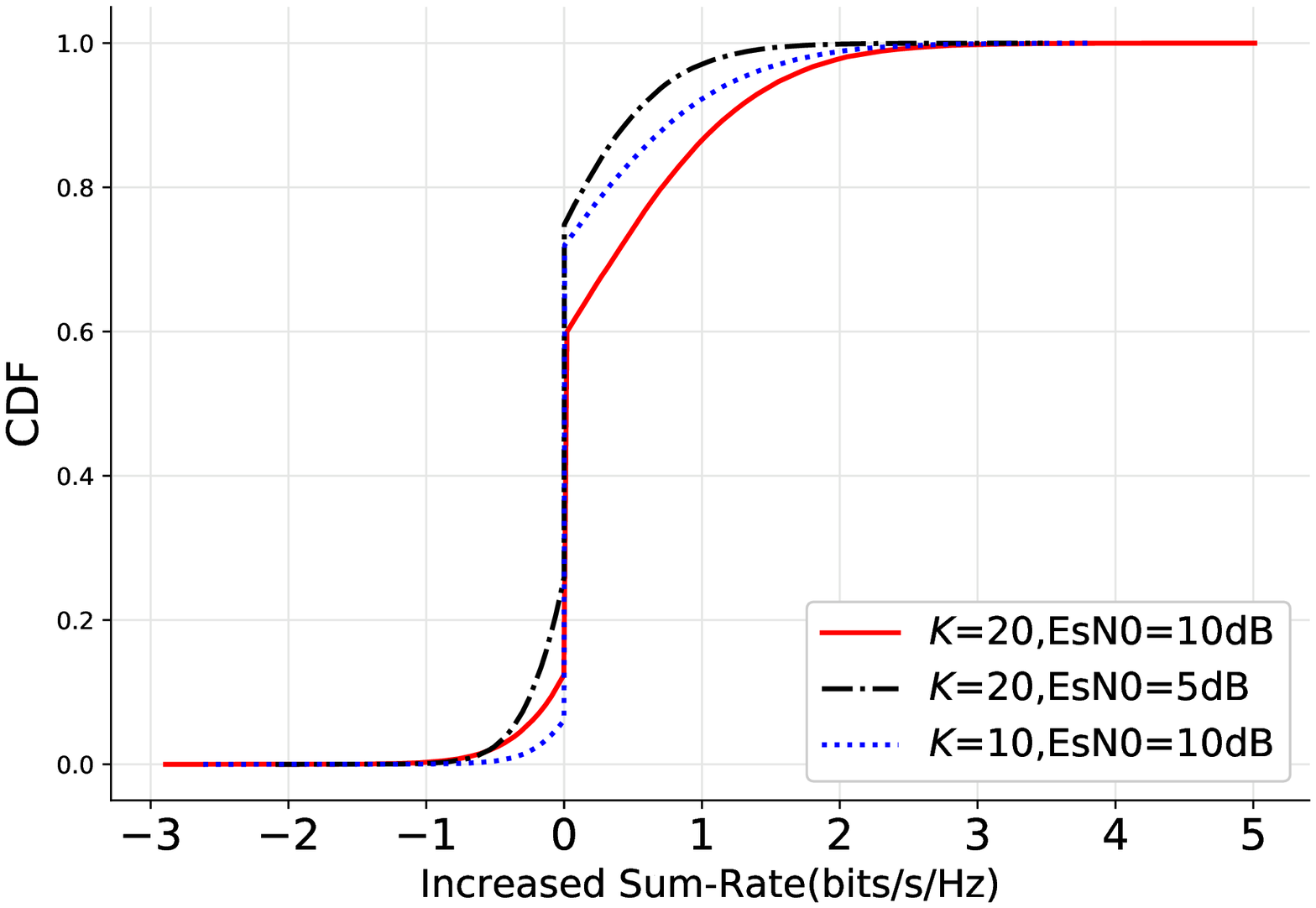}}
	\caption{Empirical CDF of $R_{PN}(\mathbf{h})-R_{c}(\mathbf{h})$,  with $M=10$ for ePCNets.}
	\label{fig:cdf_increased_rate}
\end{figure*}

Besides the average sum rate improvement, we further evaluate ePCNet by plotting CDF of $R_{PN}(\mathbf{h})-R_{c}(\mathbf{h})$ in Fig. \ref{fig:cdf_increased_rate}. The ePCNet used in this experiment has $M=10$. For WMMSE, multiple random initializations are not used. We can see from these plots that generally speaking, ePCNet cannot \emph{universally} (with respect to all channel realizations) outperform the baseline schemes. For example, compared with GBPC, ePCNet can achieve a sum-rate improvement for approximately 40\% of the channel realizations when $K=20,\ssf{EsN0}=10$dB. However, there are still about 10\% of channel realizations for which  ePCNet does not work as well as GBPC. This is not surprising because the objective of the training is to maximize the \emph{average} sum rate, and as a result ePCNet is not expected to achieve performance gain for \emph{all} network input.

\subsection{Analyzing the optimization landscape for ePCNet and WMMSE}

A direct analysis for PCNet is quite difficult due to over-parametrization in DNN. However, it has been observed and reported in the deep learning literature that for multilayer networks such as DNN, different local optima often provide similar performance, even with highly non-convex loss functions. The authors of \cite{choromanska2015loss} have investigated this phenomenon under several assumptions. We note that this is consistent with our observation in the performance of PCNet. 

To further validate the smoothness of the optimization landscape of PCNet, we empirically differentiate the contributions from local learners to the overall performance. Intuitively, if different local optimum leads to very similar sum rate performance, ensembling them would result in utilizing each local learner with \emph{approximately equal probability}. On the other hand, if significant difference exists among these local optima, different local learners would not contribute to the overall power control equally. 

\begin{figure}
	\centering
	\includegraphics[width=0.4\textwidth]{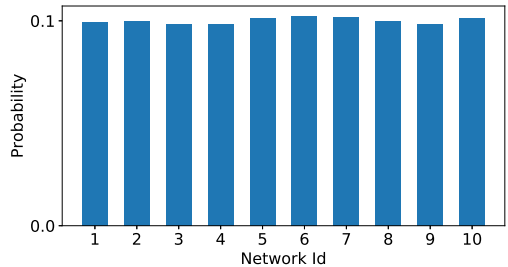}
	\caption{The histogram that each network in the ensemble is selected as the final output.}
	\label{fig:network_select_ratio}
\end{figure}

To verify this intuition, we test an ePCNet with $M=10$ and  $10^5$ samples, and plot the empirical distribution of the number of time each PCNet (i.e., local learner) is selected for the final power control. The system setting for this experiment is $K=20, \ssf{EsN0}=10$dB and the network structure is $\{400,400,200,20\}$. The ten networks are indexed from 1 to 10. Note that if more than one networks provide the (same) highest sum rate, the selected network is randomly chosen among them. The histogram is shown in Fig.~\ref{fig:network_select_ratio}. It is clear that each network in the ensemble is selected with approximately equal probability, indicating that all networks contribute similarly to the overall performance. This suggests that either majority of the networks reach the same local optimum, or they reach different ones which are very similar. Either way, the optimization landscape of PCNet is smooth.

Next we analyze the optimization landscape for WMMSE. Essentially, we want to determine whether the sum rates at different local optimal solutions for WMMSE vary significantly or not, and how this is related to the noise power.  For a specific realization of channel coefficients $\mathbf{h}$, WMMSE with random initialization will output a sum rate that is a random variable due to the random initialization, which is denoted as $r_\mathbf{h}^{(w)}$. We also denote $\mu(r_\mathbf{h}^{(w)})$ and $\sigma^2(r_\mathbf{h}^{(w)})$ as the mean and variance of $r_\mathbf{h}^{(w)}$. In the experiment, we execute the WMMSE algorithm 100 times with i.i.d. random initializations for a given channel vector $\mathbf{h}$, and obtain 100 local optimal solutions and their corresponding sum rates\footnote{Note that some of these local optimal solutions may actually be the same, which is not differentiated in the experiment.} $r_{i,\mathbf{h}}^{(w)},i=1,\cdots,100$. They can be viewed as realizations of  $r_\mathbf{h}^{(w)}$, and we can estimate the mean and variance from these realizations. 

Intuitively, $\sigma^2(r_\mathbf{h}^{(w)})$ can measure how much $r_{i,\mathbf{h}}^{(w)}$ differs from each other. In order to compare under different noise power, we also introduce a normalized measurement, coefficient of variation: 
\begin{equation}
\begin{split}
CV_\mathbf{h}=\frac{\sigma(r_\mathbf{h}^{(w)})}{\mu(r_\mathbf{h}^{(w)})}.
\end{split}
\end{equation}
Comparing the empirical distributions of  $\sigma^2(r_\mathbf{h}^{(w)})$ and $CV_\mathbf{h}$ under different noise power can help us understand the influence of background noise to the optimization landscape of WMMSE.

\begin{figure*}
	\centering
	\subfigure[Empirical CDF of variance of local optimal rates]{\includegraphics[width=0.4\textwidth]{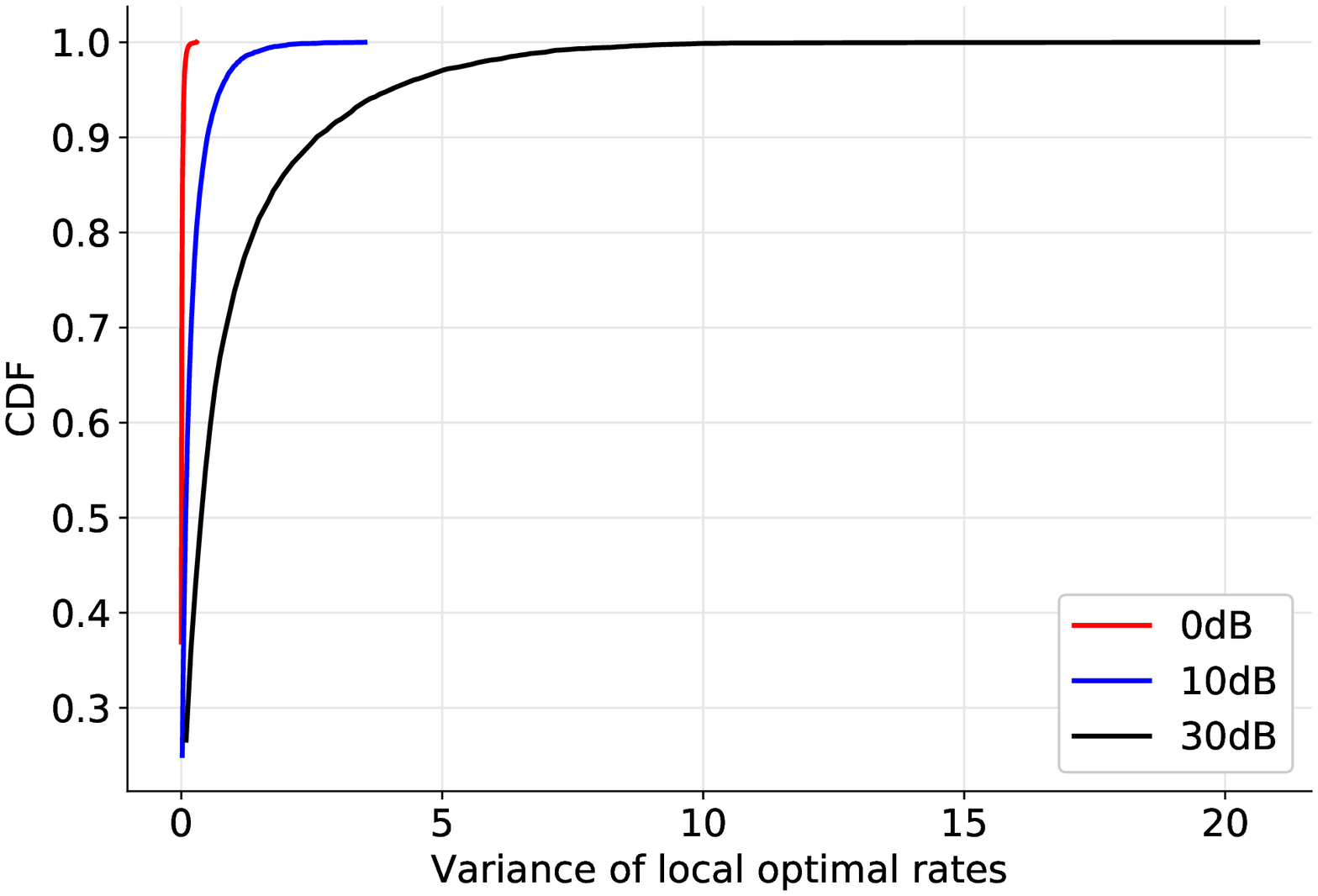} \label{fig:opt_landscape1}}
	\subfigure[Empirical CDF of coefficient of variation of local optimal rates]{\includegraphics[width=0.4\textwidth]{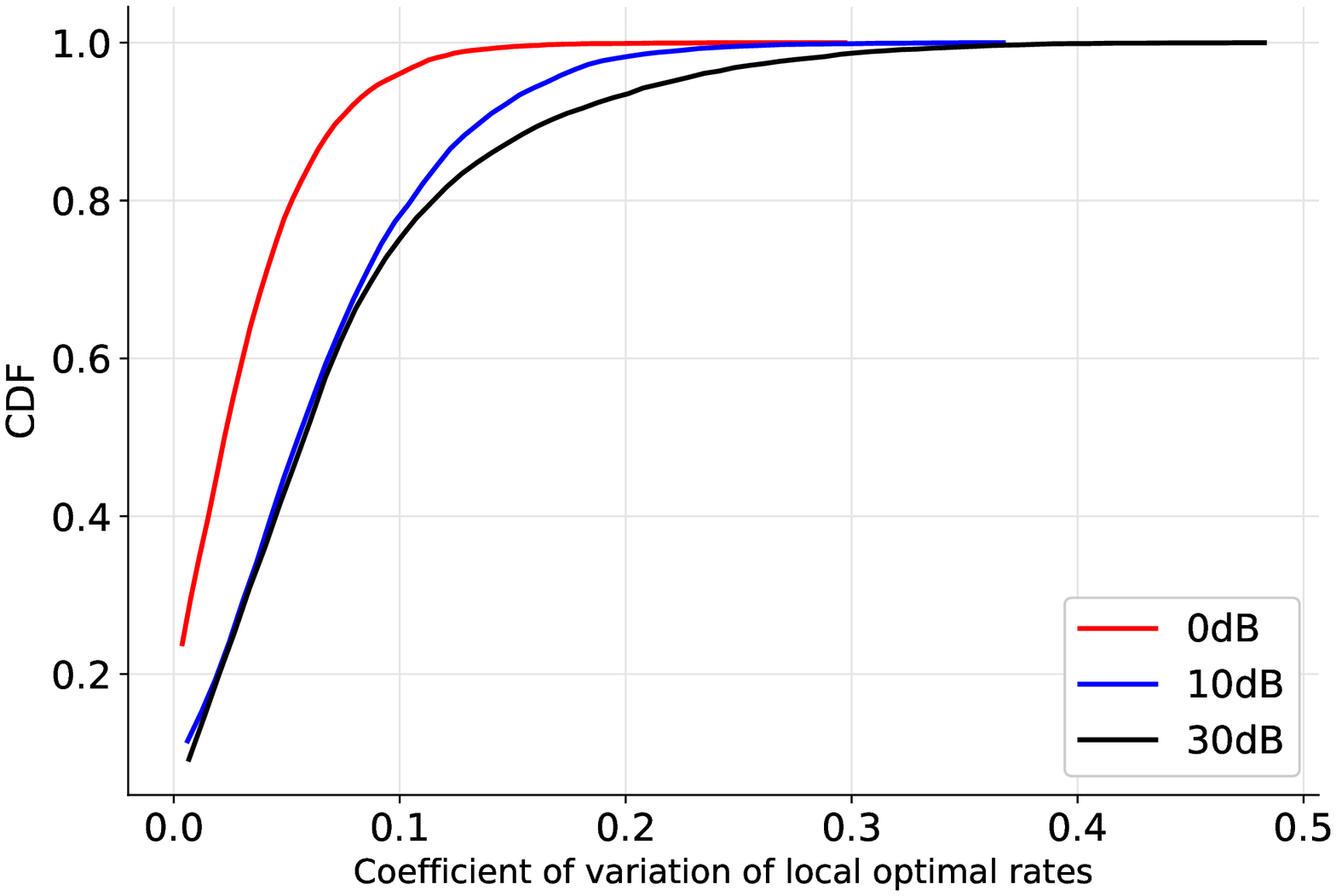} \label{fig:opt_landscape2}}
	\caption{Experimental results on how the background noise power influences the optimization landscape of WMMSE.}
	\label{fig:opt_landscape}
\end{figure*}

In Fig. \ref{fig:opt_landscape}, we plot the empirical CDFs of $\sigma^2(r_\mathbf{h}^{(w)})$ (Fig. \ref{fig:opt_landscape1}) and $CV_\mathbf{h}$ (Fig. \ref{fig:opt_landscape2}) under different values of $\ssf{EsN0}$. Each curve is obtained by simulating $10^4$ samples of $\mathbf{h}$, and for each realization of $\mathbf{h}$, 100 runs of the WMMSE algorithm with i.i.d. random initialization are carried out so that the empirical mean $\mu(r_\mathbf{h}^{(w)})$ and variance $\sigma^2(r_\mathbf{h}^{(w)})$ for this specific $\mathbf{h}$ can be computed. The empirical CDFs are then plotted with respect to the channel realizations. Both subplots show that decreasing background noise power enhances the variation of local optima of WMMSE, which may lead to bad local optimum. Combining with the result from \cite{choromanska2015loss}, this could explain why ePCNet can outperform WMMSE at high $\ssf{EsN0}$.

\subsection{Complexity analysis}

Comparing the complexity of neural networks and traditional communication algorithms is a difficult task. Simply looking at the number of floating-point operations (FLOP) is not enough for a fair conclusion, as how these operations are executed and what architecture is used also have profound impact. We note that generally PCNet has more FLOPs than WMMSE, but PCNet also has higher degree of parallelism while WMMSE is an iterative method and it has to wait for the completion of one iteration to execute the next iteration. As a result, FLOP comparison alone is not sufficient. Probably more importantly, with the development of AI technologies, neural networks have been highly optimized in some libraries such as \emph{TensorFlow} and \emph{PyTorch}. A comprehensive comparison, especially in real-world platforms, is an important topic that is worth further investigation. We also comment that although training PCNet is time and resource consuming, it only takes place in an \emph{offline} setting.

\begin{table}
\centering
	\caption{Comparison of the running time (seconds) of all considered methods.}
	\label{tab:run_time_comp}
	\begin{tabular}{l|ccccc} 
		\toprule[1pt]
		 $K, \ssf{EsN0}$ & PCNet-TF\footnotemark[2] & PCNet-NP\footnotemark[3] & WMMSE & RR & GBPC \\ 
		\midrule 
		10, 10dB & 2.7 & 1.5 & 3.8 & 261 & 28 \\ 
		20, 10dB & 3.5 & 8.4 & 5.7 & 1000 & 112 \\ 
		20, 5dB & 3.3 & 7.8 & 5.2 & 1020 & 112 \\ 
		\bottomrule 
	\end{tabular} 
\end{table}
\footnotetext[2]{PCNet-TF means PCNet is implemented with TensorFlow.}
\footnotetext[3]{PCNet-NP means PCNet is implemented with NumPy.}

In this subsection, we perform an approximate comparison in terms of the running time of all methods in the same computational environment. This may not be totally accurate but it can give us at least some qualitative complexity comparison. We implement all power control methods in Python with NumPy \cite{numpy} for algebra calculations. Specially, two version of PCNet are implemented in Python with TensorFlow \cite{abadi2016tensorflow} and NumPy respectively. All programs are run using the same Intel Core i7-6700 processor (3.40GHz). To handle the potential problem that different programs may have different CPU utilizations, we only enables a single CPU core for all simulations, i.e., we do not allow multi-core processing.

In Table \ref{tab:run_time_comp}, we report the running time of different schemes for $10^4$ channel samples. First, by comparing PCNet-TF and PCNet-NP, we find that the running time of PCNet depends on which library is used for implementation. When $K=10$, NumPy performs better while TensorFlow is more efficient when $K=20$. Compared with WMMSE, PCNet-TF has less running time but PCNet-NP runs a little slower. Overall, PCNet and WMMSE are comparable in terms of running time. In addition, RR and GBPC consume much more running time, corresponding our previous analyses.

\subsection{Performance comparison in Rician fading and different geometries}
\label{sec:sim_rician}

Although channel coefficients have been generated following the Rayleigh fading, the proposed method does not rely on any specific channel model and ePCNet can be easily applied for other channel models by retraining the PCNets. To verify this, we evaluate ePCNet under (1) a Rician fading model with 0dB $K$-factor, and (2) different pathloss values for transceiver pairs. For ePCNet, we use the same configurations as in the Rayleigh fading experiments. 

\begin{figure*}
	\centering
	\subfigure[$K=10, \ssf{EsN0}=10$dB\label{fig:rice_perf_comp_10_10}]{\includegraphics[width=0.32\textwidth]{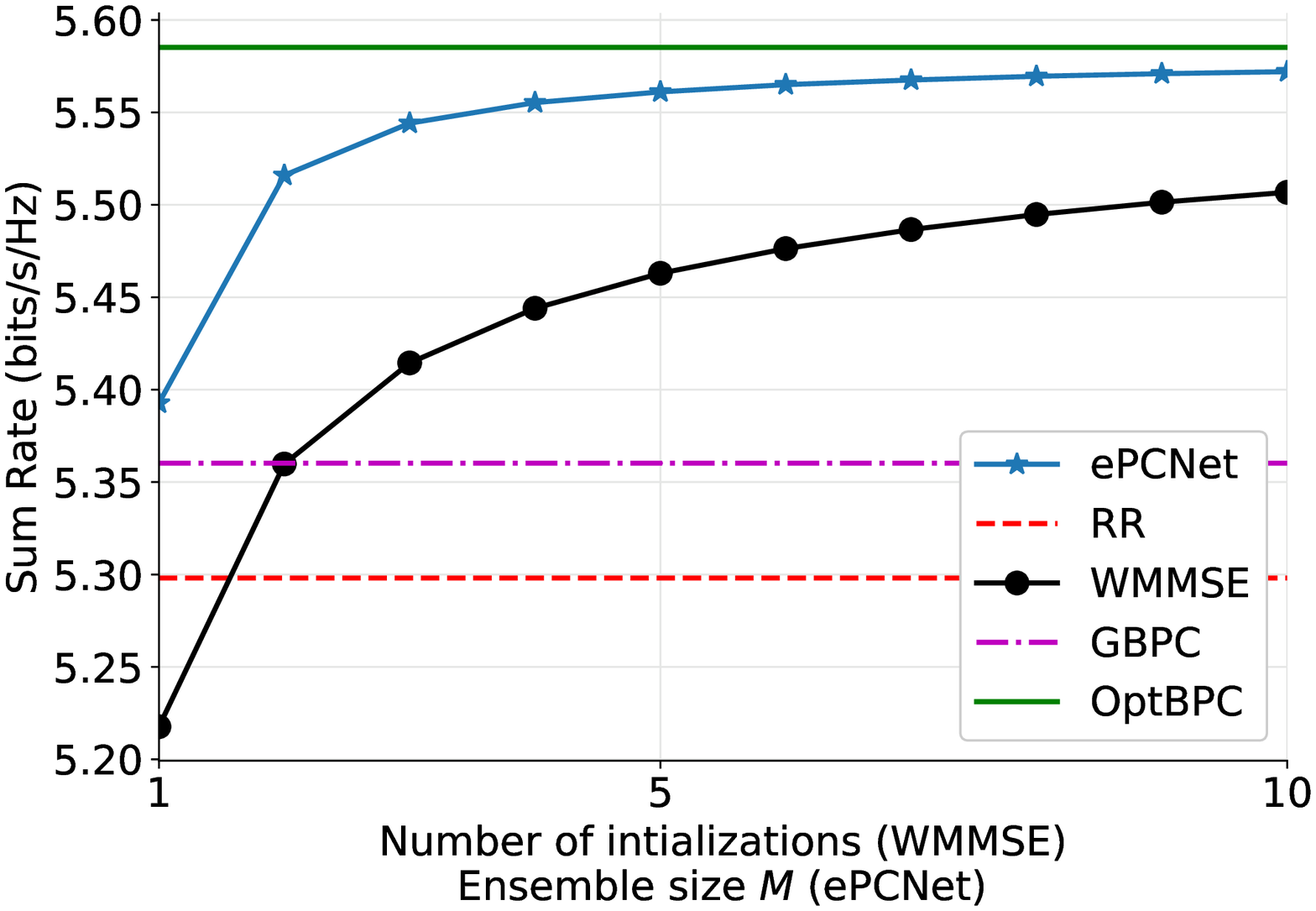}}
	\subfigure[$K=20, \ssf{EsN0}=10$dB\label{fig:rice_perf_comp_20_10}]{\includegraphics[width=0.32\textwidth]{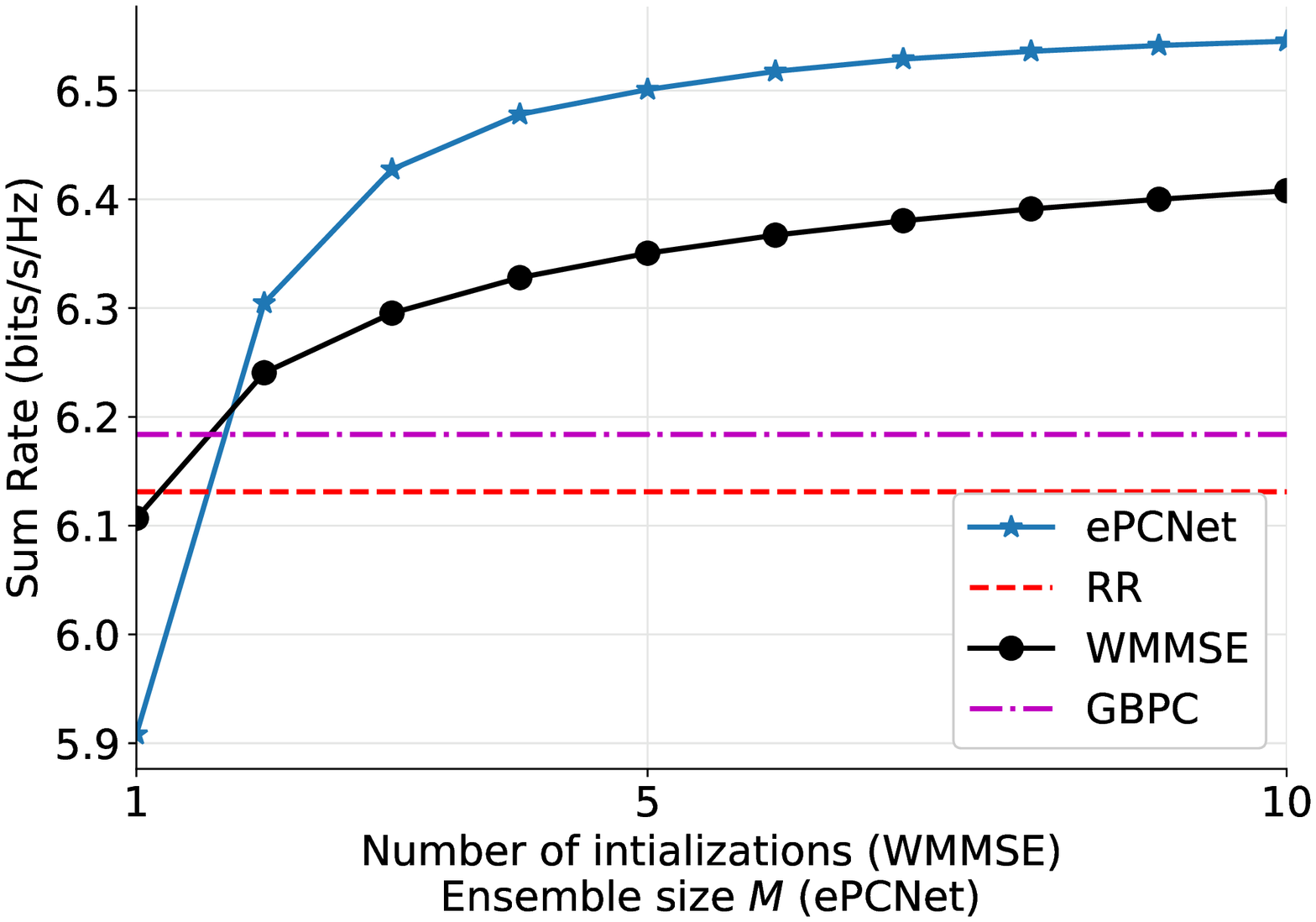}}
	\subfigure[$K=20, \ssf{EsN0}=5$dB\label{fig:rice_perf_comp_20_5}]{\includegraphics[width=0.32\textwidth]{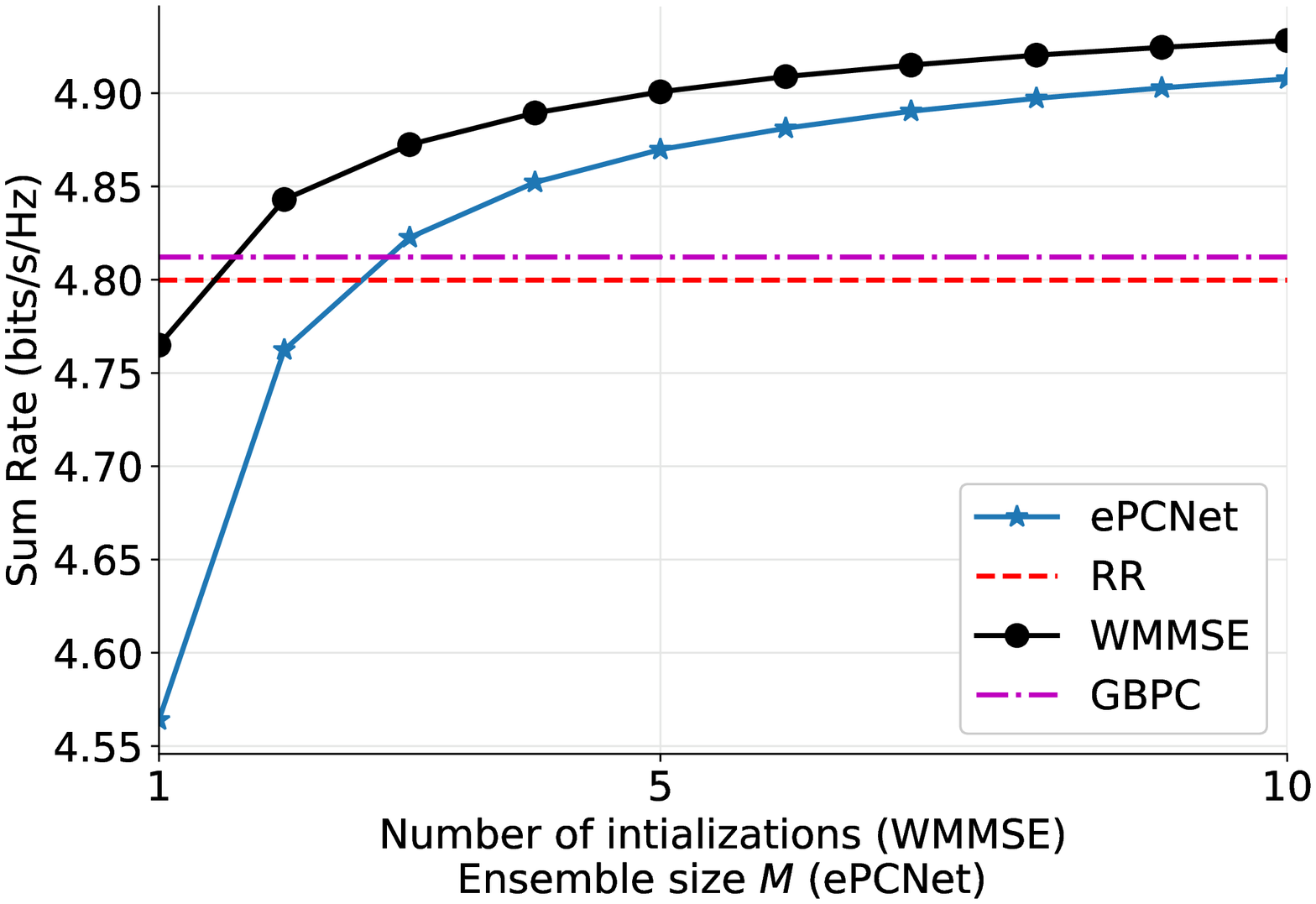}}
	\caption{Performance comparison of different schemes in Rician fading channels. The network structure is $\{100,200,100,10\}$ for $K=10$ and  $\{400,400,200,20\}$ for $K=20$.}
	\label{fig:rice_perf_comp}
\end{figure*}

The simulation results for Rician fading are given in Fig. \ref{fig:rice_perf_comp}. For WMMSE, we also give the results with multiple random initializations. It is clear that both ePCNet and WMMSE can outperform RR and GBPC by combining multiple local optima. For the comparison between ePCNet and WMMSE, ePCNet can achieve better performance with relatively low noise power. This is similar with the observations in Rayleigh channels. Therefore, we conclude that the proposed ePCNet is also effective in Rician fading channels, validating its broad applicability to different interference channel models. 

We then evaluate the performance of ePCNet by considering different geometries of transmitters and receivers. In our experiments, all transmitters and receivers are uniformly randomly distributed in a 10 meter $\times$ 10 meter area. The channel gain $\lVert h_{i,j}\rVert^2$ can be written as
$\lVert h_{i,j}\rVert^2=G_{i,j}\lVert f_{i,j}\rVert^2$,
where $f_{i,j}$ is the small-scale fading coefficient which still follows $\mathcal{CN}(0, 1)$, and $G_{i,j}$ is the pathloss function defined as \cite{inaltekin2009unbounded}
$G_{i,j}=\frac{1}{1+d_{i,j}^2}$,
where $d_{i,j}$ is the distance between the $i$-th transmitter and $j$-th receiver.

\begin{figure*}
                \centering
                \subfigure[$K=10, \ssf{EsN0}=25$dB\label{fig:geometry_perf_comp_10_25}]{\includegraphics[width=0.3\textwidth]{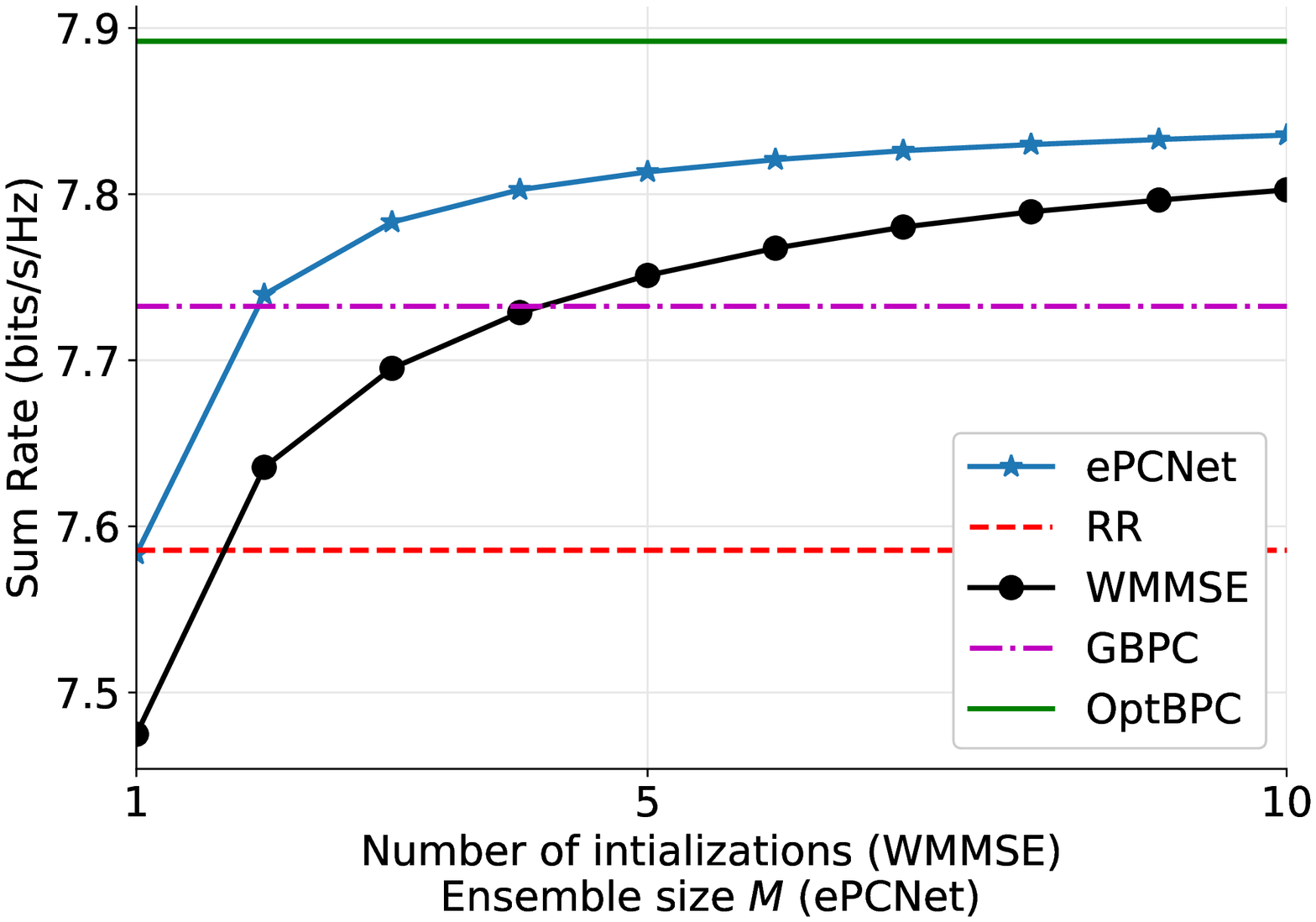}}
                \subfigure[$K=10, \ssf{EsN0}=30$dB\label{fig:geometry_perf_comp_10_30}]{\includegraphics[width=0.3\textwidth]{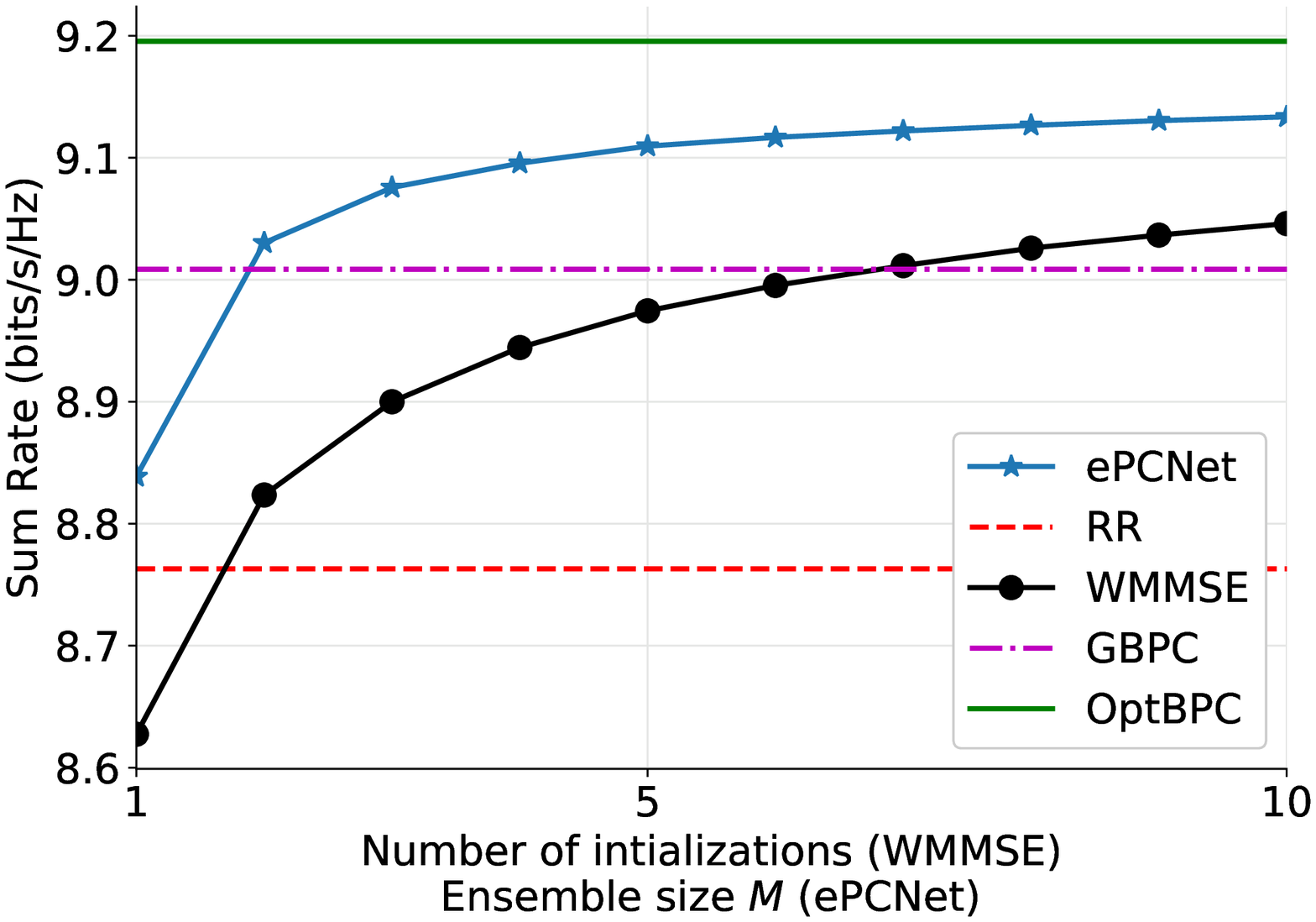}}
                \subfigure[$K=20, \ssf{EsN0}=30$dB\label{fig:geometry_perf_comp_20_30}]{\includegraphics[width=0.3\textwidth]{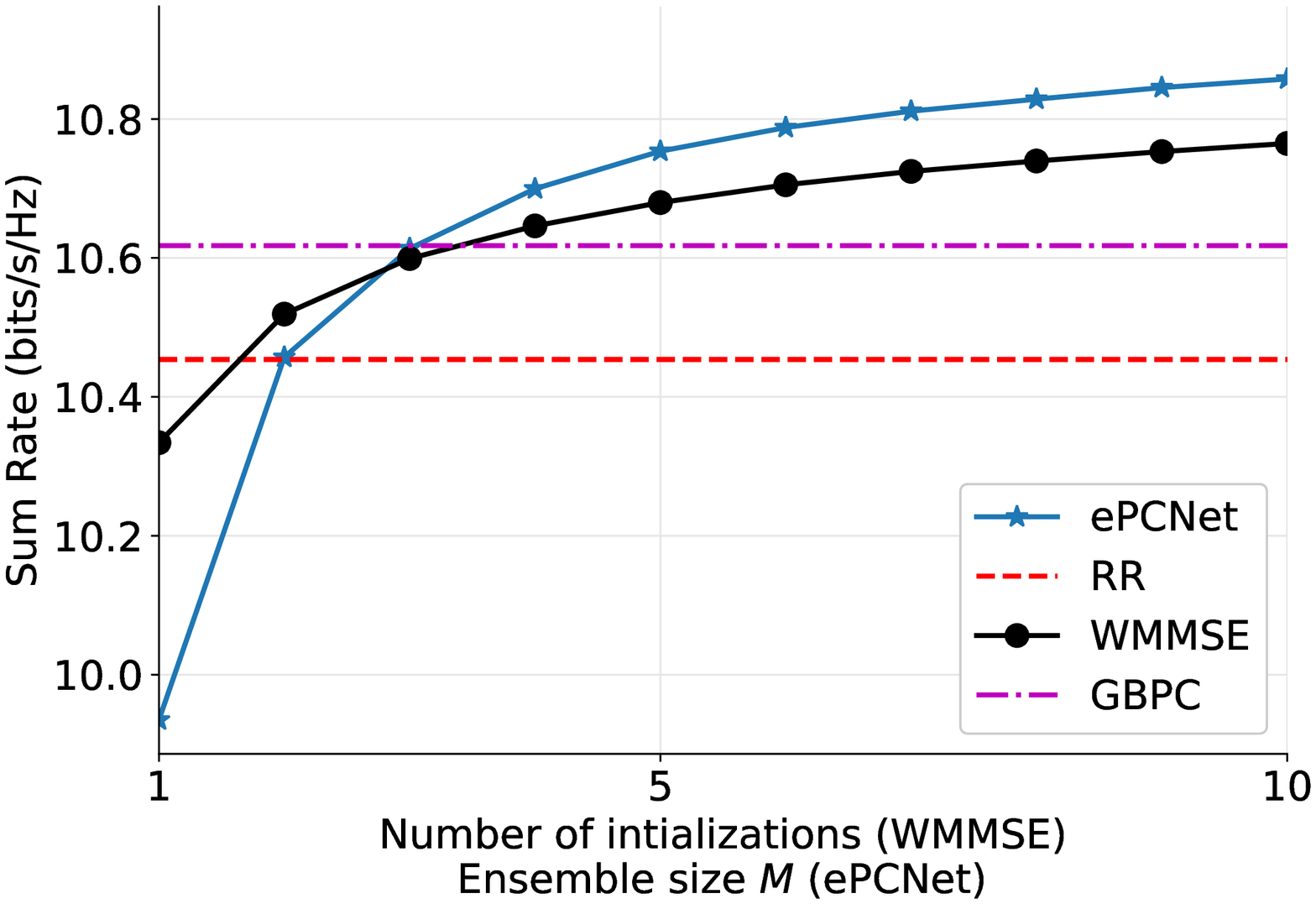}}
                \caption{Performance comparison of different schemes in different geometries. The network structure is $\{100,200,100,10\}$ for $K=10$ and $\{400,400,200,20\}$ for $K=20$.}
                \label{fig:geometry_perf_comp}
\end{figure*}

The simulation results are shown in Fig. \ref{fig:geometry_perf_comp}. It is observed that similar comparison results are obtained and ePCNet can still outperform other schemes when different geometries are modeled. This indicates again that the proposed method can be used for a wide range of channel models.

\section{Numerical Evaluations for SRM-QC}
\label{sec:simlation_srm_qc}
\subsection{Overview}
In this section, we focus on evaluating the proposed PCNet(+)/ePCNet(+) for the SRM-QC problem. The implementation details of PCNet(+)/ePCNet(+) remain the same as in the previous section 

The geometric programming based algorithm is the compared algorithm for SRM-QC. The GP-based method, which was proposed in \cite{chiang2007power}, is a well known and high-performance algorithm for SRM-QC. Specifically, we implement the GP-based method for the non-convex case with successive convex approximation. It successively approximates the original non-convex problem as several convex problems and solve these convex problems iteratively via geometric programming until the solution converges. The details of this algorithm can be found in Section IV of \cite{chiang2007power}.

Besides GP-based algorithms, SCALE \cite{papandriopoulos2009scale} one another well known method for solving the SRM-QC problem. However, this algorithm is built based on a high-signal-to-interference-ratio assumption. According to our experimental results, we find that the algorithm may not converge if this assumption is not satisfied. Therefore, we do not involve SCALE in our comparison. In addition, there still exist some methods which can find the global optimal solution, such as MAPEL \cite{qian2009mapel}. However, these algorithms find the global optimal solution based on an implicitly exhaustive search and the exponentially increasing complexity makes them unfeasible for practical usage \cite{hong2014signal}.

Due to the high complexity of GP, the results in this section are obtained by simulating $10^4$ channel samples. Unless otherwise specified, Rayleigh fading is used for simulation. Different with the evaluations in Section \ref{sec:simulation_srm}, under a specific rate constraint, we generate channel samples following a predefined distribution but only keep those which have feasible solutions for training networks and evaluations.

\subsection{Performance comparison}

\begin{figure*}
	\centering
	\includegraphics[width=0.5\linewidth]{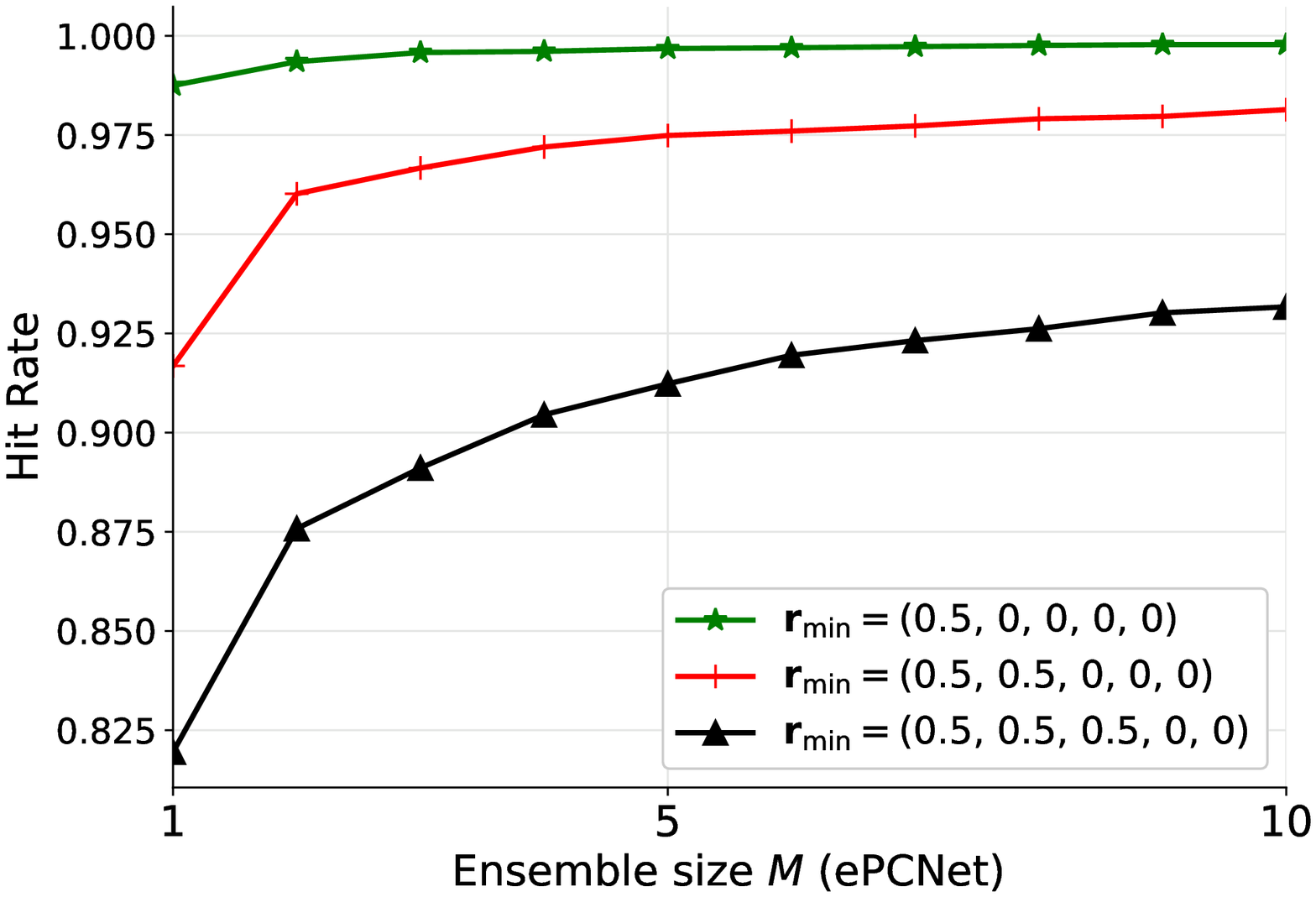}
	\caption{Hit rate of ePCNet. $K=5,\ssf{EsN0}=10$dB.}
	\label{fig:hit_rate_epcnet}
\end{figure*}

\begin{table}
	\centering
	\caption{The average sum rates of ePCNet and GP for SRM-QC.} 
	\label{tab:perf_comp_srm_qc}
	\begin{tabular}{l|cc} 
		\toprule[1pt]
		$\mathbf{r}_{\min}$ & ePCNet($M=10$) & GP \\ 
		\midrule 
		(0.5, 0, 0, 0, 0) & \bf{3.75} & 3.04\\
		(0.5, 0.5, 0, 0, 0) & \bf{3.07} & 2.87\\ 
		(0.5, 0.5, 0.5, 0, 0) & 2.86 & \bf{2.93}\\ 
		\bottomrule 
	\end{tabular} 
\end{table}

In Table \ref{tab:perf_comp_srm_qc}, we compare the achieved rates with $K=5,\ssf{EsN0}=10$dB and different minimum rate constraints. The network shape of PCNet in this case is $\{25,50,25,5\}$. The scaling factor $\lambda$ in the loss function \eqref{eqn:loss_srm_qc} is set to 10. The performance of ePCNet with $M=10$ is shown in Table \ref{tab:perf_comp_srm_qc}. It can be observed that ePCNet can achieve higher sum rate than GP when the rate constraint is not so strong. When $\mathbf{r}_{\min}=(0.5,0.5,0,0,0)$, ePCNet with $M=10$ outperforms GP by 23\%. However, when the rate constraint becomes tight, the GP-based method performs more efficiently. This is not difficult to explain. A tighter rate constraint shrinks the feasible solution space, which may contain a small number of local optima. It is more likely that GP can find the global optimal solution in such a reduced space. However, PCNet cannot perform such an efficient search because it is not tailored to the problem.

\begin{figure*}
	\centering
	\includegraphics[width=0.5\linewidth]{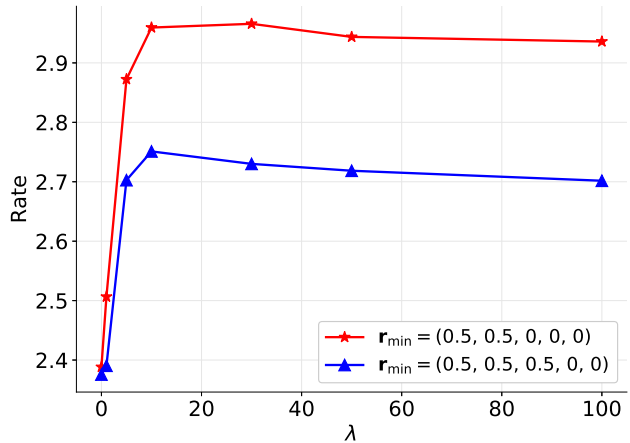}
	\caption{The influence of $\lambda$ to the performance of PCNet. $K=5,\ssf{EsN0}=10$dB.}
	\label{fig:lambda_pcnet}
\end{figure*}

A deep look at the output of ePCNet will help us better understand its behavior. We define hit rate as the ratio of the number of channel samples for which ePCNet can output a feasible solution to the number of all simulated samples. Note that all simulated samples are solvable and at least one feasible solution can be found. In Fig. \ref{fig:hit_rate_epcnet}, we show the hit rate of ePCNet under different rate constraints and ensemble sizes. First, we can find the hit rate will decrease as the rate constraint becomes stronger. When $\mathbf{r}_{\min}=(0.5,0,0,0,0)$, ePCNet can output a feasible solution nearly for all simulated samples. When $\mathbf{r}_{\min}=(0.5,0.5,0.5,0,0)$, one PCNet can only achieve a hit rate of 82\% and it can reach 94\% when ten networks is available in the ensemble. However, GP can strictly search the solution in the feasible solution space and the hit rate is undoubtedly 100\%. 

The gap of hit rate between ePCNet and GP is due to the difficulty to explicitly let the neural network know there exist some constraints to the output and the feasible solution space is reduced. PCNet cannot perform a focused optimization inside the feasible solution space, which exactly influence its efficiency when the constraint is strong.

\subsection{The impact of $\lambda$}

PCNet addresses the rate constraint by adding penalty terms to the loss function in \eqref{eqn:loss_srm_qc}, and the scaling factor $\lambda$ plays an important role in deciding the PCNet behavior. In this part, we will give some results to show how the scaling factor $\lambda$ influences the PCNet performance.

In Fig. \ref{fig:lambda_pcnet}, we show how the achieved sum rate changes of one PCNet with the scaling factor $\lambda$. It clearly shows that if we set $\lambda=0$, PCNet suffers from a severe performance degradation, indicating the necessity of introducing penalty terms in the loss function. In the future, if we can come up with some new methods to address the rate constraints, the penalty terms may be unnecessary. 
{Intuitively, $\lambda$ should be large so that the network will search in the entire feasible solution space because any violation will cause dramatic increase due to the penalty term. However, from Fig. \ref{fig:lambda_pcnet}, we find that this is not the case and selecting large $\lambda$ may  degrade the performance. The reason is that the network cannot constraint the searching space in the feasible solution space for all training samples. If the network evolves toward a direction that satisfies the constraints for some samples, the results of other samples may again violate the constraints. This means the network will keep focusing on minimizing the penalty term all the time while not paying any attention to the sum rate. Consequently, the performance may be degraded.
}

\begin{figure}
	\centering
	\includegraphics[width=0.5\linewidth]{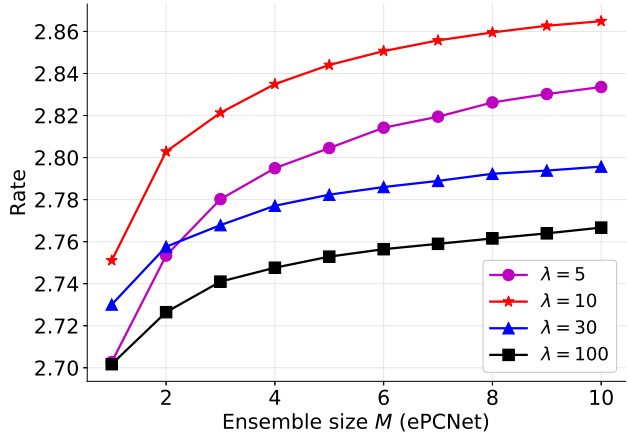}
	\caption{The influence of $\lambda$ to the performance of ePCNet. $K=5,\ssf{EsN0}=10$dB. $\mathbf{r}_{\min}=(0.5,0.5,0.5,0,0)$.}
	\label{fig:lambda_epcnet}
\end{figure}
The results in Fig. \ref{fig:lambda_epcnet} help us further understand the influence of $\lambda$. We compare the performance of ePCNet with different ensemble sizes and different $\lambda$'s. The minimum rate constraints are $\mathbf{r}_{\min}=(0.5,0.5,0.5,0,0)$. First, we can see that $\lambda=10$ not only provides better performance when only one PCNet is available, it also outperforms others with increased ensemble size. Comparing the results of $\lambda=5$ and $\lambda=30$, an interesting observation can be made. When $M=1$, $\lambda=5$ cannot outperform $\lambda=30$, but it  quickly outperforms $\lambda=30$ when the ensemble size increases. This result indicates that selecting a larger $\lambda$ may improve the sum rate by increasing the hit rate, but it may also sacrifice the ability of ePCNet to achieve a higher sum rate by combining multiple networks.

\section{Discussion and Future Works}
\label{sec:conclusion}
In this paper, we have studied a long-standing open problem from a new perspective. Inspired by recent advances in artificial intelligence, we propose using deep learning technologies to address the traditional power control problem for interference management in wireless networks. We first develop \emph{PCNet} -- a fully connected multi-layer neural network which takes the channel coefficients as input and outputs the transmit power of all transmitters.  An unsupervised learning strategy is adopted to train PCNet by directly maximizing the system sum rate. An ensemble of PCNets, i.e., \emph{ePCNet}, is proposed and shown to improve the sum-rate performance over traditional expert-based methods. By taking the noise power as an additional network input, we obtain \emph{PCNet+/ePCNet+} which provides the desirable generalization capacity. Extensive experiments have been carried out to verify the performance of ePCNet(+) and to analyze its behavior under different system settings. Sum rate and complexity comparison show that ePCNet(+) achieves better power control while consuming less computational resources. 

The key ingredient that has enabled our new solutions is leveraging training deep neural networks to solve non-convex optimization problems. Hence, the proposed framework is not limited to the two specific problems discussed in this paper. By generalizing the optimized objective and constraints, deep learning technologies have the potential to tackle other non-convex optimization problems which are often encountered in wireless communications. However, this generalization may not always be easy, as we have seen in SRM-QC versus SRM. It is difficult to optimize discrete objective functions and address general constraints with neural networks, which calls for more research  in the future.

There are some other challenges remaining to be solved to make the DNN-based power control ready for practice. First, in this work, we have shown that PCNet provides desirable generalization capacity for background noise power if it is also fed into the network. However, there are other system parameters such as the number of  users $K$ and the distribution of channel coefficients, which are not generalized. How can we generalize PCNet to incorporate these system parameters is an important research direction. Second, we currently assume ideal channel estimation, which is not realistic in practice. How to make PCNet robust against channel estimation errors should be further investigated. Last but not the least, the proposed scheme is a centralized framework. How to devise a distributed version of PCNet with little performance loss is another interesting but challenging topic.

\bibliographystyle{IEEEtran}
\bibliography{IEEEabrv}

\end{document}